\documentclass[acmsmall,table]{acmart}

\usepackage{enumitem}

\usepackage{url}

\AtBeginDocument{%
  \providecommand\BibTeX{{%
    \normalfont B\kern-0.5em{\scshape i\kern-0.25em b}\kern-0.8em\TeX}}}

\setcopyright{acmcopyright}
\copyrightyear{2024}
\acmYear{2024}
\acmBooktitle{Proceedings of The 2024 ACM SIGCHI Conference on Computer-Supported Cooperative Work \& Social Computing (CSCW 2024), November 09--13,
  2024, San Jose, Costa Rica}
\acmDOI{XXXXXXX.XXXXXXX}

\acmConference[CSCW'24]{Make sure to enter the correct
  conference title from your rights confirmation email}{November 09--13,
  2024}{San Jose, Costa Rica}
\acmPrice{15.00}
\acmISBN{978-1-4503-XXXX-X/18/06}

\newcommand{\revise}[1]{{\color{black} #1}}

\newcommand{\eg}{\emph{e.g.}}
\newcommand{\ie}{\emph{i.e.}}
\newcommand{\vs}{\emph{vs.}}

\definecolor{grey}{rgb}{0.5, 0.5, 0.5}
\newcommand{\uinput}[1]{{\color{grey}{\textours{#1}}}}

\DeclareTextFontCommand{\textours}{\fontfamily{qpl}\selectfont}

\begin{document}

\title[The Contemporary Art of Image Search]{The Contemporary Art of Image Search: Iterative User Intent Expansion via Vision-Language Model}

\author{Yilin Ye}
\email{yyebd@connect.ust.hk}
\affiliation{%
 \institution{The Hong Kong University of Science and Technology (Guangzhou)}
  \city{Guangzhou}
  \state{Guangdong}
 \country{China, }}
 \affiliation{%
 \institution{The Hong Kong University of Science and Technology}
\city{Hong Kong SAR}
 \country{China}}
\author{Qian Zhu}
\email{qian.zhu@connect.ust.hk}
 \affiliation{%
 \institution{The Hong Kong University of Science and Technology}
\city{Hong Kong SAR}
 \country{China}}
\author{Shishi Xiao}
\email{sxiao713@connect.hkust-gz.edu.cn}
\affiliation{%
 \institution{The Hong Kong University of Science and Technology (Guangzhou)}
  \city{Guangzhou}
  \state{Guangdong}
 \country{China}}
\author{Kang Zhang}
\email{kzhangcma@ust.hk}
\affiliation{%
 \institution{The Hong Kong University of Science and Technology (Guangzhou)}
  \city{Guangzhou}
  \state{Guangdong}
 \country{China, }}
 \affiliation{%
 \institution{The Hong Kong University of Science and Technology}
\city{Hong Kong SAR}
 \country{China}}
\author{Wei Zeng}
\authornote{Wei Zeng is the corresponding author}
\email{weizeng@ust.hk}
\affiliation{%
 \institution{The Hong Kong University of Science and Technology (Guangzhou)}
  \city{Guangzhou}
  \state{Guangdong}
 \country{China, }}
 \affiliation{%
 \institution{The Hong Kong University of Science and Technology}
\city{Hong Kong SAR}
 \country{China}}

\begin{abstract}
Image search is an essential and user-friendly method to explore vast galleries of digital images.
However, existing image search methods heavily rely on proximity measurements like tag matching or image similarity, requiring precise user inputs for satisfactory results.
To meet the growing demand for a contemporary image search engine that enables accurate comprehension of users' search intentions, we introduce an innovative user intent expansion framework. 
Our framework leverages visual-language models to parse and compose multi-modal user inputs to provide more accurate and satisfying results.
It comprises two-stage processes: 1) a \emph{parsing} stage that incorporates a language parsing module with large language models to enhance the comprehension of textual inputs, along with a visual parsing module that integrates an interactive segmentation module to swiftly identify detailed visual elements within images; and 2) a \emph{logic composition} stage that combines multiple user search intents into a unified logic expression for more sophisticated operations in complex searching scenarios.
Moreover, the intent expansion framework enables users to perform flexible contextualized interactions with the search results to further specify or adjust their detailed search intents iteratively. 
We implemented the framework into an image search system for NFT (non-fungible token) search and conducted a user study to evaluate its usability and novel properties.
The results indicate that the proposed framework significantly improves users' image search experience. Particularly the parsing and contextualized interactions prove useful in allowing users to express their search intents more accurately and engage in a more enjoyable iterative search experience.
\end{abstract}



\maketitle

\section{Introduction}\label{sec:Intro}
\begin{figure*}[t]
\centering
\includegraphics[width=0.995\textwidth]{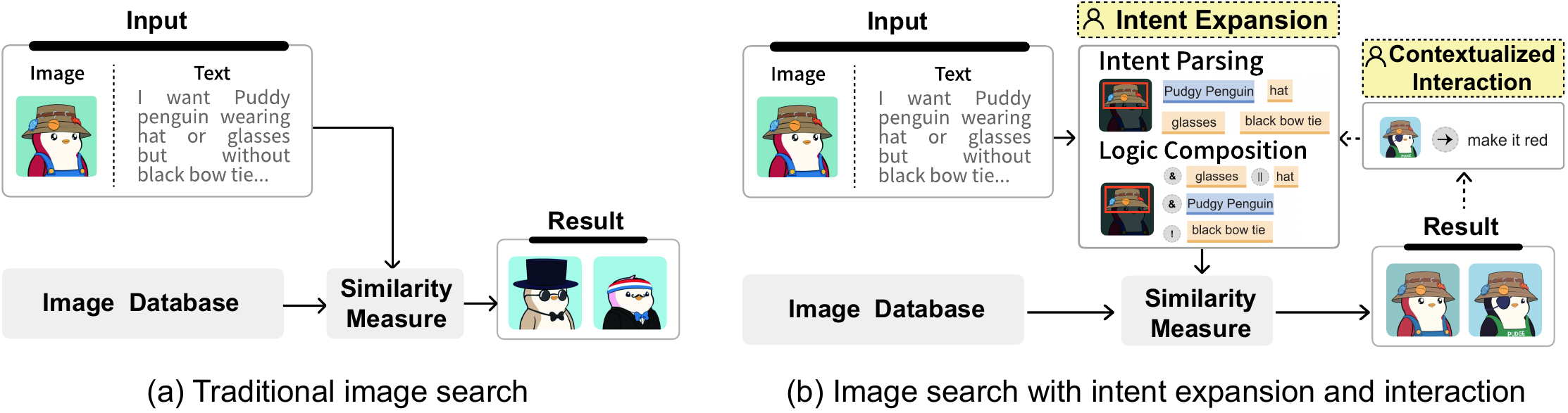}
\vspace{-2em}
\caption{Comparison of previous image search framework and our framework. (a) Previous frameworks are mostly based on proximity measurement only, which can be further divided into text-to-image search and image-to-image search, where recent development in deep learning supports more fuzzy input. However, all these methods lack sufficient comprehension of user intent before measuring the proximity and hardly support contextualized interactions with search results to iteratively adjust or specify search intents. \revise{(b) In comparison, our proposed intent expansion framework seek to interpret users' complex and logically-related multi-modal inputs and integrate into the underlying similarity measure to optimize the process. Our framework also supports enhanced contextualized cross-modal interactions to refine the search.}
}
\label{fig:framework}
\end{figure*}

Image search serves as an essential and valuable method for many purposes, such as enabling users to efficiently access paintings or designs of interest.
As such, image search plays an important role in many commercial applications~\cite{zaidi2019implementation}.
For instance, online digital art platforms (\eg, \textit{ArtStation} and \textit{DeviantArt}) allow users to understand the trend of specific styles and get inspiration for their own works~\cite{chen2019gallery,huang2019swire}.
As shown in Figure~\ref{fig:framework} (a), different image search methods can be categorized based on the input type.
Text-to-image search takes keywords as input and returns images with matching tags or descriptions~\cite{yee2003faceted, wu2012tag, li2016socializing}.
Alternatively, image-to-image search takes images or sketches as inputs and returns visually similar images~\cite{ eitz2010sketch, huang2019swire, latif2019content, dubey2021decade}.
Both methods rely on proximity measurement between user inputs and image items in a gallery, requiring precise and specific user inputs for successful retrieval of desired results.

\begin{figure*}[t]
\centering
\includegraphics[width=0.995\textwidth]{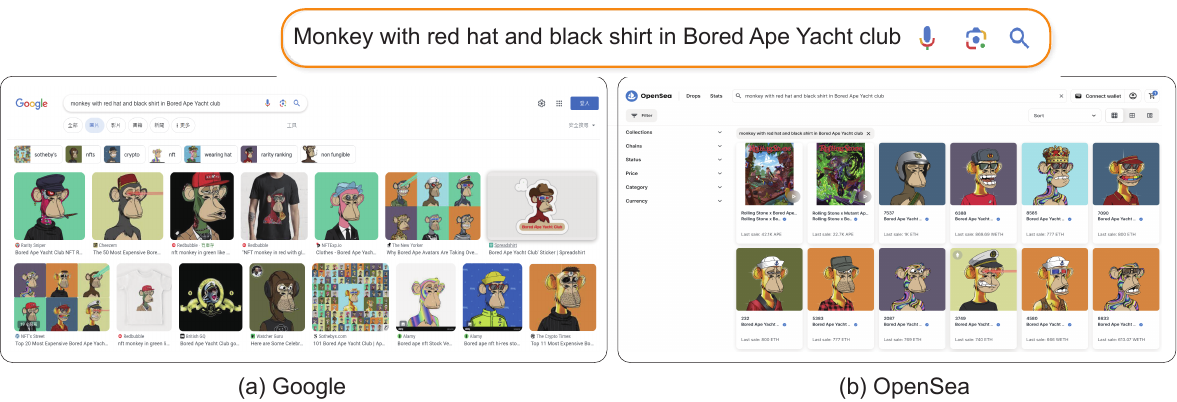}
\vspace{-2mm}
\caption{
Insufficient understanding of search intents in user input for current image search systems. When users input \uinput{"monkey with red hat and black shirt in Bored Ape Yacht club"} to (a) Google (an open-domain image search system) and (b) OpenSea (a specialized NFT image search system), they cannot get satisfactory results.
Both systems cannot understand the composed search intents of  \uinput{"Bored Ape Yacht club"}, \uinput{"red hat"} and \uinput{"black shirt"} with logical relation.
}
\label{fig:motiv}
\end{figure*}

\begin{figure*}[t]
\centering
\includegraphics[width=0.995\textwidth]{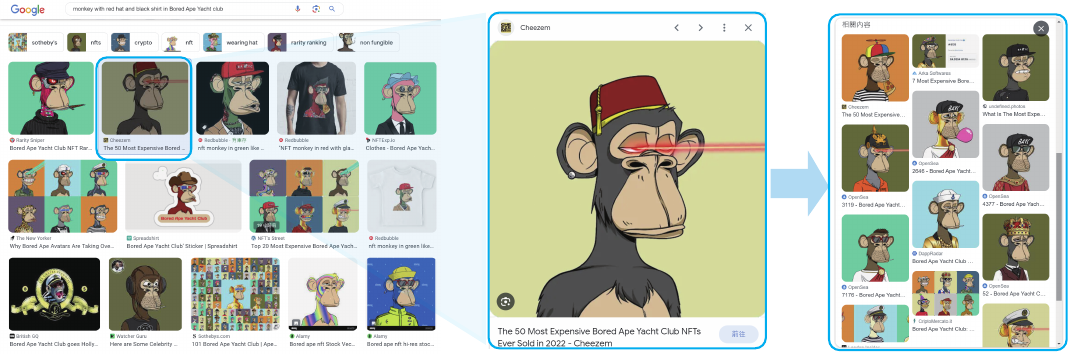}
\vspace{-2mm}
\caption{
Insufficient understanding of intents in users' feedback.
Following the example in Figure~\ref{fig:motiv}, when users use Google's interaction to click on an image, he may be interested in this special type of red hat.
However, the system's image-based iterative search only measure aggregated similarity and cannot understand such user intent.
Thus it returns images all without such hats.
Even though recently available bounding box detection function allows users to focus on particular region of the image, it lacks semantic understanding of the context and still does not support further logic.
For example, the user is unable to select both the red hat and black shirt in other images to perform more accurate image search to satisfy the initial text requirements.
}
\label{fig:motiv2}
\end{figure*}

Recently, there has been a rise in cross-modal image search methods that leverage the progress of deep learning technologies such as contrastive language image pretraining (CLIP)~\cite{radford2021learning}, enabling more flexible and natural inputs in texts or images~\cite{kuang2019fashion, wang2019camp, chen2020imram}.
For example, users can input \uinput{"a Pudgy penguin wearing a hat"}, and these methods can retrieve images containing both the visual elements of a \uinput{penguin} and a \uinput{hat}. 
Nevertheless, these methods have several limitations to be addressed.
First, they are primarily focused on capturing visual elements within images and are not suitable for understanding user intentions related to non-visual elements. 
For example, if a user requests \uinput{"a Pudgy penguin with the highest price"}, the concept of \uinput{the highest price} is not embedded in the images themselves but rather in the accompanying price metadata.
Second, these methods struggle to handle logical expressions embedded in natural languages.
For example, when a user requests
\uinput{"a Pudgy penguin wearing a hat but not glasses"}, the expectation is to retrieve images of a \uinput{penguin} with a \uinput{hat} but without \uinput{glasses}.
However, CLIP-based methods lack the ability to comprehend the logic behind \uinput{but not}, resulting in returning images of a \uinput{penguin} with both a \uinput{hat} and \uinput{glasses}.
Figure~\ref{fig:motiv} shows an example of how current image search systems poorly understand users' search intents and the logic, where both Google and OpenSea fail to capture the composed search intents of \uinput{red hat}, \uinput{black shirt} and \uinput{Bored Ape Yacht club}.
Last, to ensure an efficient and cohesive search experience, it is crucial to provide feedback mechanism based on initial results.
Such feedback mechanisms allow users to further specify or adjust their search intent interactively. 
Unfortunately, the existing search systems are equipped with limited feedback, preventing users from expressing their fine-grained requirements regarding the retrieved results.
For example, Figure~\ref{fig:motiv2} shows Google's iterative search mechanism where users can click on a result image to search for similar ones.
However, such image-based feedback cannot understand the specific visual element (the special red hat) composed with the black shirt in other images which the user really intends to focus on in subsequent searches.
This limitation hinders the ability to refine and improve the search process based on user preferences.

In light of these limitations, we conduct a preliminary user study to gather valuable insights and user requirements that will inform the development of improved image search methods.
We have identified several design goals \revise{(G1-G3)} according to user requirements \revise{(R1-R3)} (see Sect.~\ref{ssec:study} for details), \revise{including \emph{accurate parsing of cross-modal user intents} (G1), \emph{logic expressions of user intents} (G2), and \emph{contextualized cross-modal interaction} (G3).}

Achieving these goals poses a significant challenge.
With regard to \revise{design goal} G1, both text and image inputs can be inherently ambiguous, imprecise and composed of multiple intents, posing difficulties in accurately recognizing the specific user intents.
For instance, when searching for \uinput{"a Pudgy penguin with the highest price"}, the methods shall identify penguin images with price property, sort these images based on their prices in descending order, and return the first image.
Though the intention is simple, the underlying operations are rather complicated.
With regard to \revise{design goal} G2, user intents are typically conveyed using logical expressions, and even minor variations can lead to significantly distinct expectations, such as \uinput{"a Pudgy penguin wearing a hat and glasses"} \vs \uinput{"a Pudgy penguin wearing a hat but not glasses"}.
The divergence in user intents becomes more challenging to understand when bridging across different modes of input.
With respect to \revise{design goal} G3, existing image search systems do not support detailed contextualized interactions with result images as shown in Figure~\ref{fig:motiv2}.
Consequently, when users find a particular result that contains interesting or undesirable visual elements, such systems lack adequate support for capturing users' precise visual search intents due to the absence of efficient interaction mechanisms.
\revise{Even though advanced large language models can better understand diverse user intents, directly connecting LLM's output with existing search tools is just equivalent to rewriting the query, making it difficult to perform cross-modal parsing and truly convey the accurate intent elements to the search engines (design goal G1). 
Furthermore, such simple methods treating search tools as external APIs cannot adequately control the underlying search logic (design goal G2) and hardly support detailed contextualized cross-modal interaction (design goal G3).}

To address these challenges, this work introduces an innovative user intent expansion framework that leverages \revise{vision-language} models to parse and compose multi-modal user inputs.
The term \emph{"intent expansion"} means the design of search method and interaction to allow users to express their search intents in more detail.
As illustrated in Figure~\ref{fig:framework} (b), the framework mainly \revise{compromises} two components: intent parsing and intent logic composition.
First, to respond to \revise{user requirement} R1, we leverage \revise{vision-language} model to achieve accurate parsing of user inputs with both language parsing and visual parsing (Sect.~\ref{ssec:parsing}).
In language parsing, we leverage in-context few-shot learning of large language models to achieve an accurate understanding of complex user intents about visual features and metadata of images.
In visual parsing, we incorporate an interactive semantic segmentation model to allow users to specify their intents in terms of specific visual elements.
Second, the intent logic composition (\revise{user requirement} R2) leverages chain-of-thought prompting of large language models (LLMs)~\cite{Brown2020gpt} and visual interaction to recognize logic in users' intents (Sect.~\ref{ssec:composition}).
Logic-aware intent matching rules and text-guided image editing methods are employed to embed multi-modal logic composition in the retrieval process. 
Furthermore, we provide users with contextualized interaction (\revise{user requirement} R3) that combines the parsing and the logic composition to iteratively refine users' search intents (Sect.~\ref{ssec:interaction}). 

Guided by the user intent expansion framework, we develop a prototype search system (Sect.~\ref{sec:sys}) for the specific scenario of non-fungible token (NFT), recognizing its growing prominence and unique characteristics in the digital asset domain.
The prototype system contains over 20 popular collections from OpenSea$-$the biggest NFT marketplace, along with the metadata like collection name, price and tags.
The system allows cross-modal inputs (text, image and their combination) with intent parsing and logic composition and user-friendly contextualized interaction to search for NFT.
Finally, we conduct a user study to evaluate the usability and unique functions of our prototype (Sect.~\ref{sec:eval}), comparing two baseline systems of text-based metadata search and cross-modal text-to-image search without parsing. 
We find that the intent expansion and the contextualized interactions are most useful for improving users' search experience with a more detailed understanding of users' search intents and a more enjoyable iterative search experience.

Our major contributions include:
\begin{itemize}
\item 
We gather user requirements and summarize the design considerations for improving current digital image gallery search.
\item
We propose an intent expansion framework with parsing of cross-modal user intents, logic composition and contextualized cross-modal interactions.
The framework enables a more user-friendly image search with an enhanced understanding of user intents.
\item 
We develop a practical prototype system for NFT search based on our framework and conduct user experiments to demonstrate its advantages.

\end{itemize}

\section{Background and Related Work}\label{sec:related}
\textbf{Image Retrieval}.
To facilitate the exploration of large image galleries, effective retrieval methods are essential to efficiently search for the desired images.
To this end, numerous image retrieval techniques have been developed, including keyword-based search and proximity-based search. 
Keyword-based methods leverage text annotations generated by users or AI of predefined attributes~\cite{yee2003faceted, koren2008personalized,kumar2011describable} or tags~\cite{li2008learning, wu2012tag, li2016socializing}.
First, they fail to grasp users' search intent regarding image properties that are not explicitly annotated in the accompanying text.
Text annotations can only provide partial descriptions of an image, but users would not want their search to be constrained by these limited text descriptions.
Second, strict matching rules employed by these methods are insufficient to comprehend fuzzy natural language queries that contain multiple logically related elements of intent.
Users would prefer to use more natural and expressive search language instead of being confined to using specific keywords.

Proximity-based methods offer greater flexibility, allowing users to input fuzzy text and image data without the need for specific keywords.
Recent advancements in deep learning techniques have improved proximity measurement, leading to results that are semantically and visually similar~\cite{liu2016deepfashion, kuang2019fashion,lee2018stacked, wang2019camp, chen2020imram}.
Vision-language pretraining techniques~\cite{radford2021learning, jia2021scaling} have been developed to train AI models to learn the correspondence between a large number of image and text caption pairs, further enhancing the capabilities of these methods.
However, despite these advancements, the methods still face challenges in adequately understanding users' search intents.
First, natural language can be ambiguous, and the same search intent can be expressed in different ways, increasing search uncertainty and weakening the system's ability to parse cross-modal inputs and comprehend the logical relationships among multiple intents.
Moreover, the methods aggregate all visual features without being aware of the specific features that users are seeking.

Another prevalent issue with existing search methods is the lack of capability to allow users to interactively specify or adjust their search intent to iteratively refine the search results.
Previous research has explored additional interaction techniques to assist users in specifying their detailed visual search intent, including sketching~\cite{eitz2010sketch}, object extraction using example images~\cite {yeh2005picture}, concept learning based on user annotation~\cite{fogarty2008cueflik}, region-based search~\cite{huang2010review} and the lately popular composed search that incorporates users' text descriptions or AI-detected text labels~\cite{baldrati2022effective, xiao2023wytiwyr}. 
For example, composed search~\cite{baldrati2022effective} allows users to input an image along with a descriptive sentence indicating the desired changes to be made in the image before initiating the search.
However, the method often requires users to input lengthy additional text multiple times, resulting in a significant increase in user workload.
Overall, existing methods lack support for easy and intuitive visual and cross-modal interactions, making it challenging for users to interactively refine their search intent.

In our work, we first seek to address the insufficient user intent comprehension by existing image search systems with an innovative intent expansion framework.
The intent expansion consists of language and visual parsing as well as logic composition to strengthen search system's ability to capture user intents.
Second, the intent expansion further enables users to perform easy-to-use contextualized cross-modal interactions with current results, which can help them efficiently adjust their search intents in an iterative manner. 

\vspace{1.5mm}
\noindent
\textbf{Conversational Search}.
The rise of intelligent chatbots has sparked interest among HCI and CSCW researchers, giving rise to the emerging field of conversational search and conversational information seeking~\cite{zhang2018towards, vtyurina2017exploring, avula2022effects, sekulic2022evaluating}. 
Some researchers focus on conceptual aspects of conversational search, providing design frameworks and evaluation based on user simulation.
For instance, Radlinski et al.~\cite{radlinski2017theoretical} identified three important high-level requirements: 
1) users should have the ability to express their search requests using natural language without constraints; 
2) the system should be capable of suggesting search results and seeking further clarification from the user; and
3) users should have the ability to provide feedback on the results received.

Despite the rich conceptual frameworks, the implementation of a practical conversational search system is not a trivial task.
In the pursuit of incorporating conversational abilities into search systems, some researchers have taken the initiative to develop prototypes for specific applications
For instance, Zhang et al.~\cite{zhang2018towards} concentrated on using system-generated questions to clarify user requirements regarding product aspects within an e-commerce context, while Hashemi et al.~\cite{hashemi2020guided} introduced a transformer-based neural network to enhance the automatic generation of clarifying questions.
With the advent of powerful large language models (LLMs) like GPT~\cite{openai2023gpt4}, researchers and developers have started to explore the usage of LLMs in recommendation and retrieval tasks~\cite{gao2023chat, bao2023tallrec, jeronymo2023inpars, jin2023inferfix, wang2023query2doc}, along with some commercial projects like Microsoft's New Bing and Google's Bard.
However, many of the existing tools and studies are either solely focused on unimodal text documents, or rely on external retrieval applications like traditional image search engines.
As a result, the integration of multimodal capabilities, specifically combining text and images, is still an area that requires further exploration and development.
Furthermore, it is important to highlight that the majority of conversational prototypes only provide feedback interaction in single-modality texts.
This can be cumbersome as users need to repeatedly input text to refine their search, and it is challenging to accurately describe certain visual features due to the inherent disparity between the text and image modalities.
There is a need for more sophisticated approaches that bridge the gap between text and image modalities, allowing users to provide more intuitive and efficient feedback during the conversational image search process.

Our work seeks to integrate more user-friendly interactive feedback into the cross-modal image search.   
Particularly, we propose contextualized cross-modal interactions to facilitate more flexible and sophisticated feedback beyond text.

\vspace{1.5mm}
\noindent
\textbf{User Prompting of Vision-Language Model}.
The recent development of large pretrained language and vision AI models have significantly enhance their ability to understand fuzzy input by users, mainly through different user prompts.
Such models have shown great potential for improving the ability to capture user intents in diverse inputs for HCI applications.

The most popular form of prompt for these models is text.
LLMs have shown superior ability in understanding \revise{users' diverse text inputs}.
With special prompting techniques like chain-of-thought ~\cite{wei2022cot, wang2022iteratively} and tree-of-thought~\cite{yao2023tree}, LLMs can perform accurate reasoning~\cite{chan2022data} and convert users' fuzzy input into precise code or structured query understandable by computer programs.
This makes it possible to develop robust language interface to connect with many traditional applications to streamline human computer interaction.
For example, TaskMatrixAI~\cite{liang2023taskmatrix} defines a general framework for the ecosystem of connecting LLM to numerous external APIs, which consists of four modules: 1) multimodal conversational foundation model, 2) API platform, 3) API selector, and 4) API executor.
HuggingGPT~\cite{shen2023hugginggpt} connects user inputs and requirements for an AI task expressed in natural language to the most suitable HuggingFace model to produce the results.
Some vision-language models also accept text prompts for cross-modal tasks, such as text-to-image generation~\cite{rombach2022high, xiao2023let} and text-guided image editing~\cite{brooks2023instructpix2pix}.

Some researchers also explore visual prompt for vision and language models, because in some cases language alone cannot sufficiently describe users' detailed requirements, especially in visual or cross-modal tasks, such as visual question answering~\cite{li2023blip} and interactive image segmentation~\cite{kirillov2023segment, li2023semantic}.
For example, the Segment Anything Model~\cite{kirillov2023segment} (SAM) allows for both text and visual prompts on the user's part to segment fine-grained elements in an image.
The visual prompts of SAM support intuitive clicking on any position in the image and easy box selection of semantic regions.
The enhanced comprehension of users' visual input is starting to show great potential in HCI tasks involving vision data, such as user-customized image inpainting~\cite{yu2023inpaint}.

However, how such prompting can be exploited to optimize image search experience still remains a question due to challenges like the large volume of data concerned in retrieval and the insufficient cross-modal alignment. 
\revise{More importantly, it is not enough to directly connect state-of-the-art LLM with existing image search systems, because even though the LLM can interpret users' intentions, the interpretation still need to be input to the search systems in plain text, which cannot optimize the underlying search logic and interaction process.}
Our work leverages both textual and visual prompting of vision and language models combined with \revise{optimization of the underlying search including} cross-modal alignment \revise{and logic-aware intent matching and ranking rules} to boost the image search system's ability to grasp user intents in their multi-modal input.

\section{Formative Study}\label{sec:Framework}
We conduct a formative study to understand the current image search practices with target users, identify the potential requirements they have with existing systems, and summarize the corresponding design goals of image search systems for NFT (non-fungible token) search.

\subsection{Study Setup and Procedure}\label{ssec:study}

\noindent
\textbf{Participants}:
We recruited seven participants aged from 23 to 33 (three females) by personal connection in our university. Three of them have a design background and the other four are majoring in computational art.
All the participants are familiar with image search platforms and frequently used image search to find designs or artworks online for appreciation or inspiration.
All participants have known about NFT prior to the study.

\noindent
\textbf{Procedure}:
With the institute's IRB approval, we collected the participants' demographic and introduced the three existing online image search platforms to them (i.e., \textit{OpenSea}\footnote{\url{https://opensea.io/}}, \textit{Google}\footnote{\url{https://www.google.com/}} and \textit{Numbers}\footnote{\url{https://nftsearch.site/}}). These platforms are widely used image search systems based on the literature~\cite{yee2003faceted, wu2012tag, chen2020imram}.
Next, we let the participants freely search on these systems for around five minutes to get familiar with them in case any of them have not used the selected platforms.
After that, we gave participants a concrete targeted search task by randomly providing one NFT image sampled from five popular collections (i.e., \textit{Bored Ape Yacht Club, Cryptopunk, Cool Cat, Doodles, and the Doge Pound}) along with its metadata and trait tags.
The participants were first only given the image without other information.
Only after they try searching without success, we provided them with extra information about metadata.
This is because in some systems like \textit{OpenSea}, it can be difficult to find any similar results without knowing metadata.
Based on the information, the participant needed to find the 10 most similar NFTs they personally preferred.
In the process, we asked the participants to think aloud and collected their feedback by audio recording and taking notes.
After that, we conducted semi-structured interviews to ask about the difficulties they may have with the existing image search, followed by questions about their potential requirements for the image search systems.
Based on the participants' qualitative feedback, we coded the data and summarized three key requirements based on existing image search systems.

\subsection{User Requirements}
Two authors analyzed and coded the user feedback based on the audio recordings using open coding~\cite{charmaz2006constructing}. Then, the first author discussed with two other co-authors to revise and verify the themes obtained. 
Finally, we summarized three user requirements as follows.

\textbf{R1: The image search system needs to accurately comprehend the nuanced intentions of users.}
%%%%%%%%%%%%%%%
All participants in the study emphasized the necessity of an image search tool capable of understanding their inputs, including detailed properties. This need arises from their search intentions for NFT images, which often involve specific features like clothing, accessories, or background color. However, the three prevalent NFT image search systems failed to provide support for such fine-grained searches.
Participants in the study observed that \textit{OpenSea} strictly matches their textual input with the image titles, resulting in limited retrieval if the input does not precisely match the image labels. This forces users to input strictly accurate words and makes them feel that the system fails to understand their intentions correctly. 

On the other hand, while \textit{Google} offers more flexibility in text-matching, it heavily relies on textual descriptions or metadata, often leading to confusion as the input text may not explicitly match the image results. 
For instance, when a participant searched for \uinput{``monkey wearing hat''}, \textit{Google} may return an advertisement featuring a hat adorned with a monkey's face, which deviated from their intended expectation.
Furthermore, participants pointed out that \textit{Numbers} provides more precise results by considering text-visual matching. However, they found it primarily focuses on this aspect and neglects other valuable metadata, such as NFT prices. 
Moreover, the \textit{Numbers} occasionally returns confusing results despite clear and precise text input, highlighting its heavy reliance on text-visual mapping. 
For example, when P6 input a short phrase \uinput{"monkey in Bored Ape Yacht club"} which contains the collection keyword \uinput{"Bored Ape Yacht club"}, the system could not recognize this specific word correctly and may return images containing monkey but not in the target collection. Moreover, the system even misunderstood the word \emph{"Yacht"} as describing the image content and return some images with a boat (yacht). 

\textbf{R2: The image retrieval systems should be able to comprehend the logical relationships between different keywords input by users.}
During the study, four participants mentioned that they often implicitly express logical connections between multiple properties of the desired NFT image in their input, using operators such as \emph{AND}, \emph{OR}, or even \emph{NOT}. For example, P2 used the input \uinput{"monkey with a red hat AND black shirt"} to search for NFTs.
However, four participants reported that the existing image search systems (i.e., \textit{OpenSea}, \textit{Google}, and \textit{Numbers}) failed to correctly interpret such logical expressions in their input. These systems typically ignore the logical relationships and solely consider the individual input properties. As a result, for the input \uinput{"monkey with a red hat AND black shirt"}, existing systems would only retrieve images of hats, monkeys or shirts, rather than images depicting precisely a monkey wearing a red hat and a black shirt.

\textbf{R3: The image retrieval systems need to allow users to easily and efficiently provide feedback on the retrieved images}.
Four participants (P2-P5) expressed difficulties in describing very specific visual details in their search queries, resulting in difficulty in finding images with fully satisfactory visual features among the search results.
For instance, P3 searched for \uinput{"Doodles NFT with long hair"} and discovered an image with the desired hairstyle. However, when using Google's interaction of clicking on the image to find similar ones, P3 found that most of the ``similar'' images returned by Google did not possess the specific hair she was seeking.
Participants expressed the need for an efficient method that allows them to refine their search input beyond traditional text-based interactions. They found the current text-based interaction, such as tag filtering in Google, to be limited. 
Unfortunately, existing image-based input methods do not support users in providing personalized or detailed feedback on local features of the retrieved images. Participants could only click on an image to examine its details or view the recommended similar images.

\subsection{Design Goals}
Based on the participants' feedback, we distill the following design goals for improving user experience in using an image search system.

\begin{itemize}

\item
\textbf{G1: Providing Accurate Parsing of User Search Intents}. 
To satisfy \revise{user requirement} R1,  we need to provide accurate parsing of user intents in their input texts. Using multi-modal parsing is a suitable way because based on the formative study, we found that people may express their search intents with different modalities when searching NFTs, including the visual aspects embedded in the images and the metadata, such as NFT collection name and price, associated with the search images.
Therefore, it is essential to leverage multi-modal parsing modules in the image retrieval systems to comprehend users' search intents of images accurately.

\item
\textbf{G2: Enabling the Understanding of the Users' Input Logic Expressions}.
According to \revise{user requirement} R2, it is necessary to offer a logic composition module to enhance the understanding of logic expressions in users' search intents, including the commonly used logic expressions, such as \emph{"and"}, \emph{"or"} and \emph{"not"}. 
In addition, considering the NFT search or other images that may contain fine-grained features required by users, the system could also enable users' further revision of their input logic expressions associated with the retrieved visual images.

\item
\textbf{G3: Offering contextualized cross-modal Interaction}.
Based on the participants' feedback (\revise{user requirement} R3), the image search system needs to add interactive mechanisms that allow users to specify their intent based on the search results.
Users may focus on detailed visual elements and express further search requirements within an iterative process. 
Such interactions can be combined with current result images to facilitate feedback with enhanced user intent comprehension using cross-modal parsing.

\end{itemize}

\section{Framework Overview}\label{sec:overview}

%%%%%%%%%%%%%%%%%%%%%%%%%%%%%%%%%%%%%%%%%%%%%%%%%%%%%
% \subsection{VisioLingua Explorer Framework}
On the basis of the design goals, we propose a novel image retrieval framework with intent expansion that can seamlessly recognize the multi-modal and logically-related user intents. Moreover, it allows users to easily give feedback on the search results by selecting semantic regions in the images to further express their fine-grained intents.
The intent expansion framework consists of two major modules: intent parsing and intent logic composition.
The intent parsing includes language parsing and visual parsing to recognize users' search intents expressed by textual or visual input.
Based on the results of intent parsing, the intent composition module achieves composition of multiple intent elements with different logical relations expressed by users.

\revise{
Our framework is different from and more complex than directly connecting an LLM to existing image search tools like Google. 
This is because no matter how powerful the state-of-the-art LLM is (\eg, GPT-4V~\cite{yang2023dawn}), it is just a generative model in itself, which can only translate users' text or image input into another piece of text with a modified expression of users' intentions.
However, in this way, such LLM generated queries still need to be input into the existing black-box search tools as usual,  which cannot affect how the search tools understand complex user intentions. 
Essentially, this means that the LLM is just helping users rewrite the text query without actually improving the underlying search process itself.
For example, if the search engine like Google does not understand the negative logic in \emph{"woman without black hair"}, no matter how the LLM translates the initial input, (\emph{"woman, no black hair"}, \emph{"positive: woman, negative: black hair"}, etc.), when these "interpreted" inputs are fed into Google, the search results still do not match the intentions.
Alternatively, if the LLM interprets the query as "woman with blonde hair", this will change the meaning of the initial query.
In comparison, our method deeply embeds the multi-modal intentions parsed by vision-language model into the backend retrieval algorithm, explicitly and reliably guiding it to understand logical composition of multi-modal inputs. 
Specifically, our intent parsing produces a machine comprehensible data representation of users' intentions, which cannot be understood by existing image search tools but can be used to modify the backend retrieval mechanism as shown in Section~\ref{sec:sys}. 
}

\begin{figure*}[t]
\centering
\includegraphics[width=0.995\textwidth]{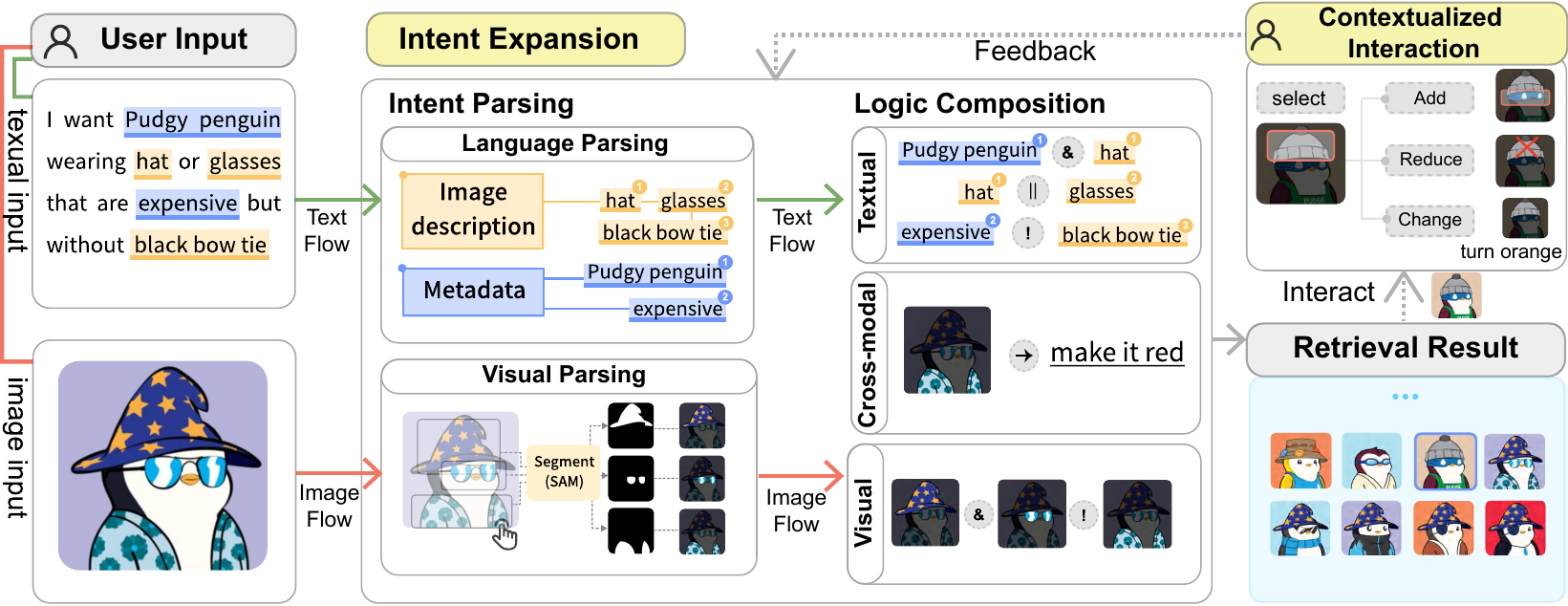}
\vspace{-2mm}
\caption{
Overview of our framework. Our framework leverages parsing-based \emph{intent expansion} to improve comprehension of users' search intents.
The intent expansion consists of two major modules: \emph{intent parsing} and \emph{logic composition}.
The intent parsing comprises both language parsing and visual parsing to capture detailed search intents of different modalities.
The logic composition can further grasp the logical relations among users' search intents to \revise{compose complex multi-element intention}.
In addition, our intent expansion provides novel contextualized interactions that allow users to directly operate on result images to iteratively refine their search.   
}
\label{fig:overview}
\end{figure*}

\subsection{Intent Parsing}
\label{ssec:parsing}
The intent parsing enhances comprehension of user intents in both language and visual input.
\subsubsection{Language Parsing for Textual Intent Recognition}
As shown in Figure~\ref{fig:overview}, we first design language parsing to recognize the user intent expressed in text.
To recognize the complex intent in user inputs, we construct a domain-specific user intent taxonomy that includes the metadata, the trait tags for each NFT, and the image content description.
For the metadata, we list the two most important attributes of NFT: collection and price while more attributes can be flexibly added based on the available metadata in the system. 
Our language parsing can detect the elements in user input that correspond to the metadata.
Next, other elements in the text input may correspond to either more general visual content or the specific trait tags annotated by other users.
The language parsing model detects the elements in user input and separately matches them with the images and the tags.

To implement language parsing, first we seek to automatically recognize different kinds of user intent in free-form text queries and produce a structured intermediate representation of user intent for subsequent operations.
We leverage the strong in-context learning~\cite{chan2022data} ability of LLM which can enable the model to learn how to perform a new task from few-shot examples without being specially trained for it.  
Each intent element is represented as a string element that can be organized in lists or other structures in subsequent operations to reflect logic.
We set the default intent element to be about visual content, such as \emph{``dog''}, \emph{``blue eyes''}, etc.
For metadata, we use special prefixes for each attribute.
For example, if the user wants the Azuki collection, which is a popular NFT collection featuring anime-style drawing, the recognized intent element will be \emph{``C\_Azuki''}.

\subsubsection{Visual Parsing for Visual Intent Recognition}
As shown in Figure~\ref{fig:overview}, we also design the visual parsing module to allow users to refine their intent based on visual interaction with the search results.
Users can perform easy brushing over the intended element in the image to quickly select it and search for images with similar visual elements.

For implementation of visual parsing,  we leverage an intelligent segmentation method that can more efficiently recognize user intent for particular visual element.
To be specific, we integrate the interactive Segment Anything Model (SAM)~\cite{kirillov2023segment} into the search.
SAM has two major advantages.
First, it is pretrained on large amount of semantic segmentation data in the general domain and thus is equipped with strong understanding of semantics.
Second, unlike totally automatic segmentation by machine, SAM allows users to provide visual cues for the segmentation.
One of the easiest-to-use visual cues is the box selection, where a box covering the majority of a visual element can prompt the model to locate the exact element.
This enables us to design efficient rectangle brushing to help users parse the elements before performing subsequent search.

\begin{figure*}[t]
\centering
\includegraphics[width=0.995\textwidth]{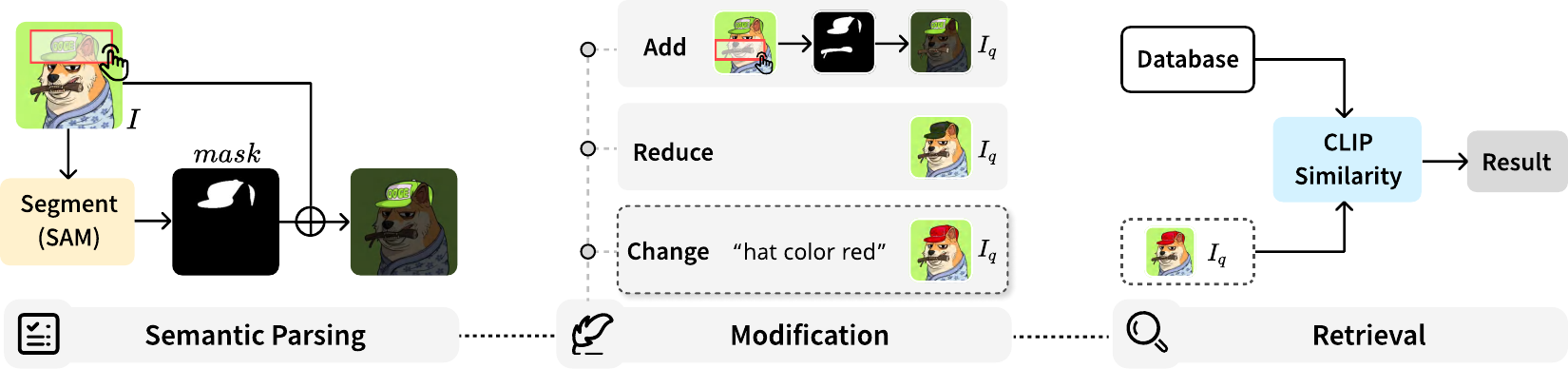}
\vspace{-2mm}
\caption{Our Visual Parsing module leverages an interactive semantic parsing method based on the Segment Anything Model, which allows users to focus on particular visual elements by simple brushing on input image $I$. Then an optional modification step allows users to express retrieval logic on the selected elements.
Finally, the query $I_q$ processed by the visual parsing is fed to database to retrieve images according to CLIP similarity.}
\label{fig:vis_parse}
\end{figure*}

In particular, the visual parsing and search is integrated as shown in Figure~\ref{fig:vis_parse}.
First, the user brushes out a rectangle over the intended element.
The positions of the four corners of the rectangle along with the image itself are passed to SAM, which produces a binary mask where the pixel value of one represents the pixel belonging to the selected visual element.
Then, the mask is applied to the image to produce two masked images, one with black pixels outside the selected element ($mask_b(I)$) and the other with white pixels outside ($mask_w(I)$).
The black masked image is combined with the original image in a weighted average ($mask_r(I)$) for regularization to avoid the influence of large black region.
\vspace{1mm}
\begin{gather}
mask_r(I)=\alpha_{0}*mask_b(I)+\alpha_{1}*I,
\end{gather}
where $\alpha_{0}=0.9$ and $\alpha_{1}=0.1$.

Next, the representation for the retrieval is computed as the average of the CLIP embeddings of $mask_r(I)$ and $mask_w(I)$.

\begin{figure*}[t]
\centering
\includegraphics[width=0.995\textwidth]{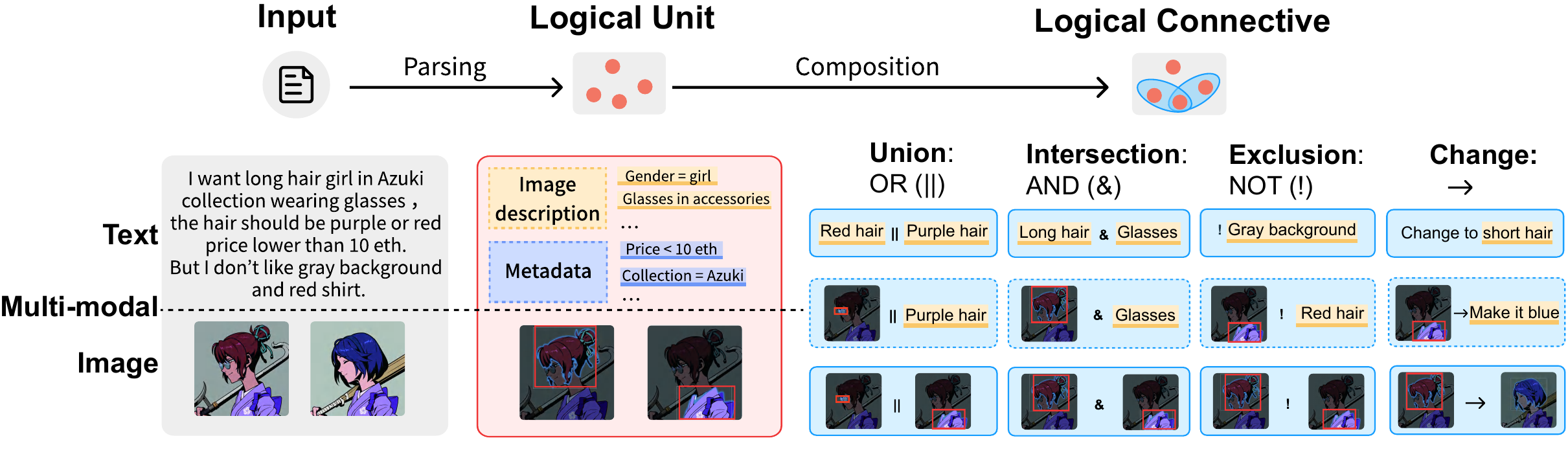}
\vspace{-2mm}
\caption{The logical composition in our framework includes four major logical connectives: union, intersection, exclusion and change, where union denotes the \emph{OR} relation between different alternative intents, intersection denotes the \emph{AND} relation between two co-ocurring intent elements, and exclusion denotes the \emph{NOT} relation to exclude unwanted elements, and the change relation to describe modification.}
\label{fig:logic}
\end{figure*}

\subsection{Intent Logic Composition}
\label{ssec:composition}
After recognizing different types of user intent, we get one or multiple intent elements.
% Users could express multiple intent units in a free-form search input.
As shown in Figure~\ref{fig:logic}, the different intent elements are combined into a composed user intent through four major logical connectives: \emph{union}, \emph{intersection}, \emph{exclusion} and \emph{change}.
The union connective signifies the user's intention to include different alternatives in the search results.
The intersection means users express multiple requirements that the search results must satisfy.
The exclusion connective represents users' negative attitudes towards certain features they don't want in the results.
The change connective adds additional input to specify the expected change of certain property of the result images. 

\subsubsection{Textual Logic Composition}

\begin{figure*}[t]
\centering
\includegraphics[width=0.995\textwidth]{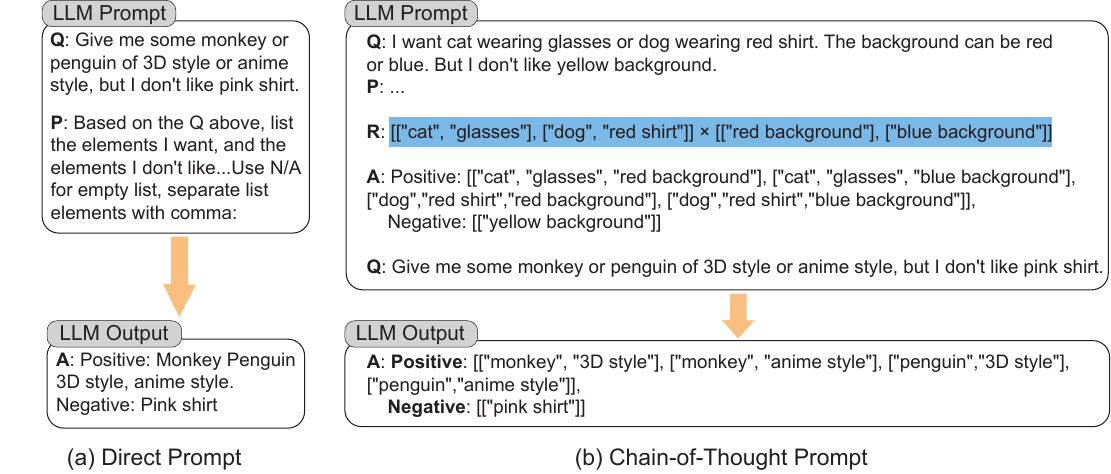}
\vspace{-2mm}
\caption{Chain-of-Thought (COT) Prompting. The COT prompting technique can enhance the reasoning ability to large language models to understand the logic expressed in users' natural language and produce more stable structured output compared to direct prompting. \revise{Specifically, as shown in (a), the query (Q) and prompt (P) stating the requirements are not enough. To improve LLM's understanding, as in (b) we provide full in-context examples containing the query (Q), the requirement prompt (P), the intermediate reasoning (R), and structured example answer (A).}}
\label{fig:cot}
\end{figure*}
We adopt special prompting technique to make sure LLM can understand the logic expressed by users in fuzzy natural language, 
aiming to reduce the system's uncertainty and unexpected mistakes caused by AI's misunderstanding of simple instruction prompt. 
Specifically, we adopt the chain-of-thought prompting~\cite{wei2022cot, wang2022iteratively}, which refers to a special prompting method that not only specifies the questions and requirements on the output but also provides exemplary intermediate human-like reasoning steps for the LLM to imitate.
Figure~\ref{fig:cot} illustrates how we achieve accurate logical reasoning with Chain-of-Thought prompting. 
Specifically, in Figure~\ref{fig:cot} (a) we can see that the direct prompting can extract the key elements but cannot clarify the logical relations.
In contrast, as shown in Figure~\ref{fig:cot} (b), we represent the logical reasoning as a nested list structure where each element represents a logical unit as in Figure~\ref{fig:logic}.
Elements separated by a comma in an inner list represent the \emph{intersection} relation, while the \emph{union} relation is signified by the comma in the outer list.
In addition, the \emph{exclusion} relation is handled by a separate list of negative elements.
Likewise, the \emph{change} relation is encoded into another list.
We then explicitly provide examples of set operations (Cartesian product) to show the LLM how to derive the results.

\subsubsection{Visual Logic Composition}
The visual logic is mainly expressed by users through easy interaction.
As shown in the optional modification step in Figure~\ref{fig:overview}, users can express logic in the visual interaction where they can select multiple elements that co-occur or express negative intent about the selected element to reduce it from the search result.
Similarly, as shown in Figure~\ref{fig:logic}, we support four types of composition relations as in text composition.

Figure~\ref{fig:vis_parse} shows more detail of this process.
First, users can perform add operation to select multiple elements and then retrieve similar images with intersection or union of the elements according to their choice.
Second, users can express their negative intent about the selected element and perform the exclusion to reduce it from the results.
Third, users can provide another example image and specify particular elements they want to change to as shown in Figure~\ref{fig:logic} and Figure~\ref{fig:vis_parse}.
After these optional operations, we obtain the query $I_q$ and measure similarity with the images in the database as shown in Figure~\ref{fig:vis_parse}.

\subsubsection{Cross-modal Logic Composition}
Furthermore, users can even express logic composition between text and visual elements, as shown in Figure~\ref{fig:logic}.
The available logical relations are similar to those in text composition and visual composition.
Particularly, in the change option, we not only allow users to enter text description but also provide a preview function to enable users to quickly see the effects of the modification on the image as shown in Figure~\ref{fig:vis_parse}.
Specifically, we combine the mask generated by SAM with InstructPix2Pix~\cite{brooks2023instructpix2pix}, a fast text-guided image editing model based on finetuned Stable Diffusion.
InstructPix2Pix takes as input the current image along with the user's modification intent and generates an edited image.
Then, the SAM mask is added to the edited image to extract the part corresponding to the previously selected element, which is then swapped with the original element.
In this way, we provide more control on the editing and make sure it only affects the selected element.

\subsection{Contextualized Cross-modal Interaction}
\label{ssec:interaction}
As shown in Figure~\ref{fig:overview}, the intent parsing and logic composition of our framework can be combined to facilitate contextualized interaction with search results.
This means that when users see the search results, sometimes they would like to continue their search based on the context of current results instead of starting from scratch.
Users may find that most of the results are not satisfactory, but some of them contain specific visual elements they like.
Similarly, due to some errors in image features extraction, they may also find some results with unexpected visual elements which might be interesting or annoying to them.
For this purpose, users can leverage our parsing and cross-modal interactions to directly operate on result images to specify their iterative search intent.
As shown in Figure~\ref{fig:overview} and Figure~\ref{fig:vis_parse}, with visual parsing users can first select specific visual elements they want to focus on.
Then they can interactively specify their intents to add, reduce or change visual elements with both visual and textual input, which is supported by our logic composition as shown in Figure~\ref{fig:logic}.

Specifically, users first can click on an image in the search results, and the system will display a pop-up window as shown in Figure~\ref{fig:interface} (c).
In this window, users can directly interact with the image to select particular visual elements.
Users can also upload other example images from local storage and interact with the newly uploaded image in the same way, which helps them specify more diverse combination of visual elements.
In \revise{case} users are not satisfied with expressing their fine-grained intents only in visuals, they can easily input a short piece of text to describe the extra features they want, with intersection, union, exclusion or change as shown in Figure~\ref{fig:logic}.
For example, the intersection allows users' feedback to describe additional features they want to add to the search results.
The exclusion allows user to input simple text to remove more general features that cannot be easily identified by visual parsing, such as blood which is drawn as numerous small drops in the image.
The change relation allows users' feedback to change certain features in the results while keeping other features similar.

\section{System Design and Implementation}\label{sec:sys}
Based on our intent expansion framework, we build a prototype for the usage scenario of NFT image search.  
In this section, we explain in detail how our search prototype is built, including the data, the similarity measurement and ranking rule implementation, and general description of our interface.

\subsection{NFT Data Collection}
To test the effectiveness and user experience of the proposed framework in a practical domain, we first collect NFT image data online.
Specifically, we gather NFT images with the corresponding metadata and textual tags from OpenSea\footnote{The biggest online NFT marketplace in the world.}.
The metadata includes contract address, token id, image path, chain, collection and price.
The tags include a set of user-labeled properties about the content or other features which vary across different collections.
In total, we obtain a large gallery and build a NFT dataset consisting of more than 500,000 NFT images.

\subsection{Multi-modal Similarity Measurement}\label{ssec:sim}
\begin{figure*}[t]
\centering
\includegraphics[width=0.995\textwidth]{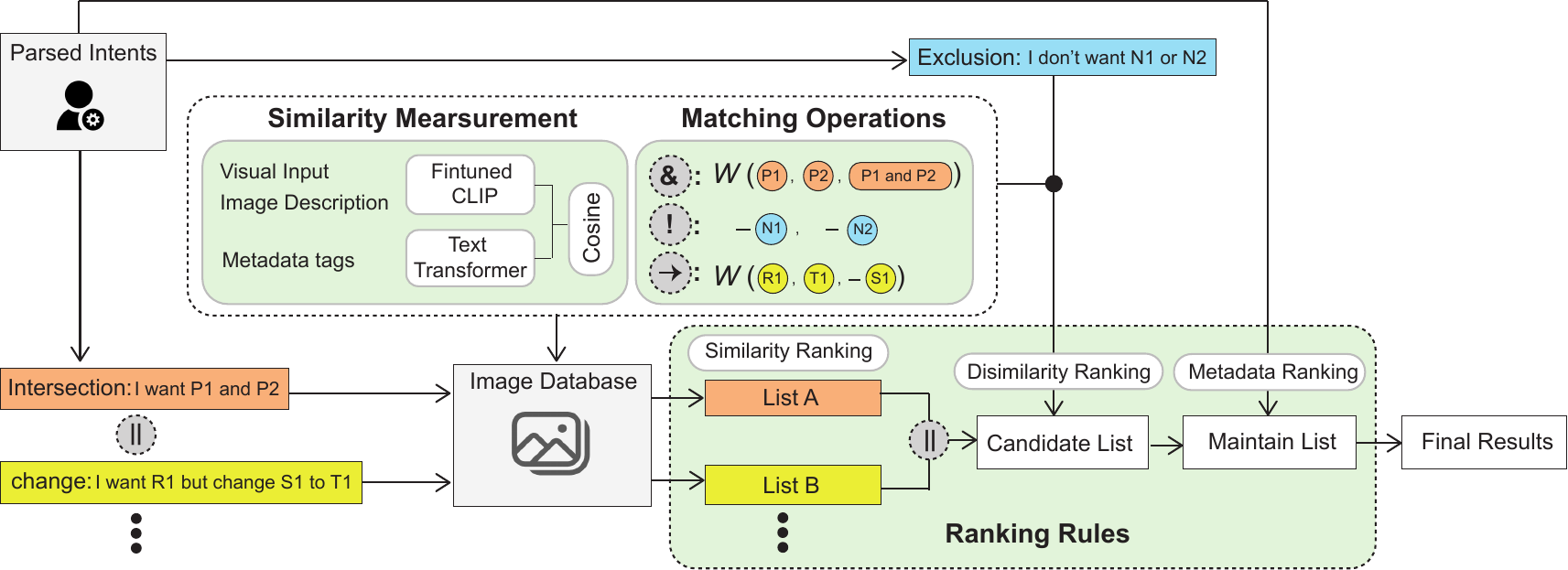}
\vspace{-2mm}
\caption{\revise{Overall workflow of our system's similarity measurement, intent matching and ranking rules. Based on the parsed intents, users' complex search can be divided into multiple logically related search intents, such as intersecting positive elements (P1 and P2), excluding negative elements (N1 and N2), changing elements from source (S1) to target (T1).
Then, the multi-modal intents are encoded into high dimensional vectors with CLIP and text transformer, and the basic similarity is measured by cosine distance.
Subsequently, the intents go through matching operations such as weighted sum (W) and minus (-) and undergo a series of ranking operations to produce final results.
}
}
\label{fig:match}
\end{figure*}

After our parsing for enhanced user intent comprehension, to incorporate text input into image retrieval, we still need to match multi-modal semantics and measure the similarity between text and image, \revise{as shown in Figure~\ref{fig:match}}.
To this end, we adopt the pretrained CLIP multi-modal embedding model~\cite{radford2021learning} to extract semantic features of both text and images into high-dimensional latent space vectors.
CLIP (Contrastive Language-Image Pretraining) is a pretrained vision-language model that jointly learns semantic representations of text and images in a common feature embedding space.
It has proven to be effective in capturing multi-modal semantics and has been successfully applied to various applications like image captioning~\cite{hessel2021clipscore}, sketch-image semantic matching~\cite{vinker2022clipasso}, image emotion alignment~\cite{wang2023reprompt} and text conditioned image generation as in the famous OpenAI DALL-E2~\cite{ramesh2022hierarchical} and Stable Diffusion~\cite{rombach2022high}.
The CLIP model acts as an encoder to extract image features and text features into implicit representation vectors. 
We then measure the \textit{Cosine distance} in the latent space to represent text-image similarity, which is a common metric for semantic alignment in high dimensional non-linear neural embedding spaces like CLIP:
\vspace{1mm}
\begin{gather}
D(t,i)=1-cos(CLIP(t),\ CLIP(i)),
\end{gather}
where $t$ and $i$ denote text and image respectively.

However, as the CLIP model is pretrained on a large number of natural images, it requires some domain transfer to adapt to the particular styles of NFT for better semantic alignment.
We achieve this by introducing few-shot contrastive finetuning on NFT domain data, as constrastive learning has proven to be effective in enhancing the embedding representation~\cite{radford2021learning, xia2023con}.
Specifically, we first utilize CLIP with the pretrained weights to retrieve adversarial examples where CLIP fails to capture some of the visual content.
Then we combine these adversarial examples with positive examples in a few-shot finetuning scheme to correct its errors.
For example, to improve the retrieval performance for input with logical relations as detected by the LLM, we provide few-shot examples to enhance CLIP's understanding of intersection, which is helpful for the rule-based intent matching in Section~\ref{ssec:rule}.
Another example is to improve CLIP's ability to distinguish between the visual features of foreground and background.
We use the triplet loss function with a threshold to train the model to distinguish between correct and incorrect results~\cite{schroff2015facenet}.
\vspace{1mm}
\begin{gather}
\| Sim(x_q, x_p) -1 \|-\| Sim(x_q, x_a) -1 \|+\alpha,
\end{gather}
where $Sim(.\ ,\ .)$ denotes the cosine similarity in the high dimensional CLIP embedding space (512 dimensions), and $\alpha$ is set to $0.05$.

Although CLIP performs well in measuring semantic similarity between text and image, the CLIP representations of the query and data items require computation of distance in high dimensional space for K nearest neighbors, which may involve high time complexity as the number of images in the gallery grows.
To reduce time complexity and improve the retrieval speed, we structure the CLIP index of the gallery into a Ball Tree~\cite{moore2003new}, where query time grows approximately as $O(D\cdot log(N))$, $D$ denotes the dimensions and $N$ denotes a total number of images in the gallery.
The Ball Tree construction and query require the distance metric to satisfy the properties of Euclidean distance, which is not the case for our original cosine distance. 
To adapt the cosine distance to Ball Tree, we normalize all CLIP embedding vectors to unit length and exploit the property that Euclidean distance ranking on the unit sphere in high dimensional space is equivalent to cosine distance ranking. 
\vspace{1mm}
\begin{gather}
cosD(u,v)\leq cosD(u,w) \Leftrightarrow \|\frac{u}{\|u\|}-\frac{v}{\|v\|}\|_2 \leq \|\frac{u}{\|u\|}-\frac{w}{\|w\|}\|_2.
\end{gather}

Similarly, the image-to-image search after visual parsing is also based on CLIP features.
In addition, for text similarity between the elements in user input and the metadata and tags, we use a pretrained text transformer to extract embeddings of semantics and measure the similarity between text embeddings with cosine distance.
The implementation of the distance metrics is by Python with the Pytorch and Sklearn packages.

\subsection{Intent Matching and Ranking Rules}\label{ssec:rule}
With the recognized intent units, their logical relations, and the similarity measure, the next step is to determine how to rank the results with combination of rules and high-dimensional similarity, \revise{as shown in Figure~\ref{fig:match}}.
Because of the existence of multiple intent elements and the logical relations, the ranking rules of the results need to be designed accordingly.
For this purpose, we regard each inner list in the nested list in Figure~\ref{fig:cot} as a composed positive intent with one or more intent elements that should appear together, and different inner lists represent different options.
Therefore, we first separately use each inner list to retrieve images.
As illustrated in Section~\ref{ssec:sim}, we already break down the task of matching complex user intent into matching with multiple elements of intersection relation and finetune the CLIP measure to enhance its understanding of intersection.
To ensure more accurate retrieved results, we combine individual element CLIP similarity with the composed intent CLIP similarity through filtering and weighted average to strengthen the intersection constraint.
Specifically, we first filter out the top 500 results with the composed intent similarity.
Then we recompute the ranking based on the weighted average of the composed intent similarity and all the single element similarity.
\vspace{1mm}
\begin{gather}
w\cdot Sim(I,v)+\sum_{i=1}^{m}w_i\cdot Sim(I[i],v),
\end{gather}
where $w$ is empirically set to 1 and all the $w_i$ set to 0.5.

Subsequently, to reinforce the intersection relation, we set a threshold to filter out all the candidates with lower than average similarity to any single element.
Similarly, for the change relation, we retrieve results according to the weighted average of similarity to the original input and the reference target element users want to change to, along with the dissimilarity to the element users intend to change. 
Next, for different composed intents of union relation, we combine the results by iteratively adding one result image from the results of each composed intent to the candidate list.
Then, for user intent of the exclusion relation, we create a maintain list by performing another ranking operation according to the similarity to the negative intent, sorted by descending order.
We remove the bottom 40\% of the results from the maintain list.
Only the images in the maintain list can be kept in the results.
If the user expresses additional ranking preferences or filtering range like price, the final list is reordered or filtered according the price requirement.

\begin{figure*}[t]
\centering
\includegraphics[width=0.995\textwidth]{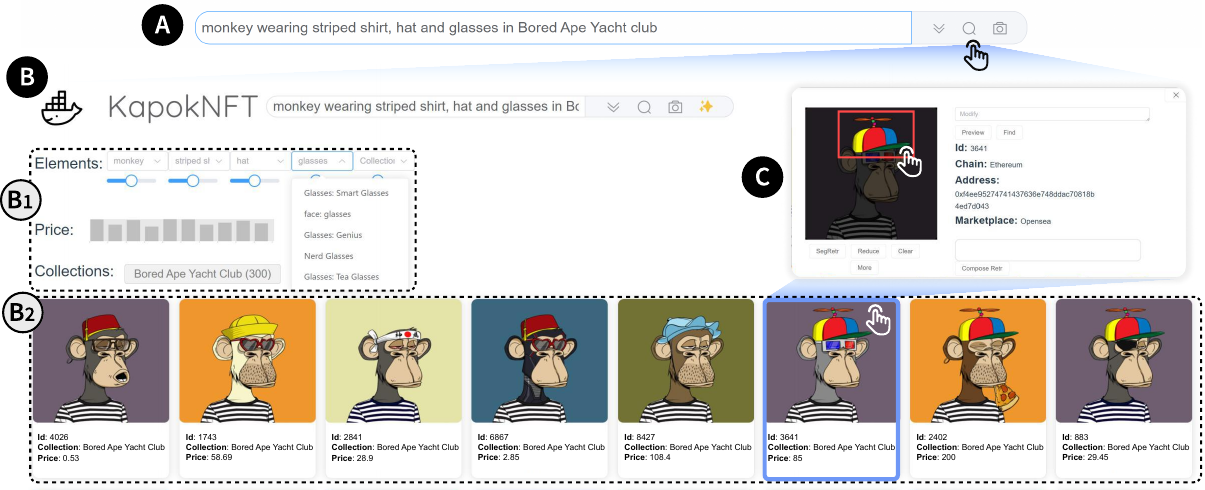}
\vspace{-2mm}
\caption{The simple interface consists of (a) an entry page, (b) a result page and (c) a pop-up window to show detail of a specific result image.
The entry page allows for both text and image input.
The result page shows result images and intent elements recognized by language parsing.
The pop-up window shows detail information and allows visual parsing interaction.}
\label{fig:interface}
\end{figure*}

\subsection{Visual Interface}

The visual interface is shown in Figure~\ref{fig:interface}.
It is a simple search prototype that consists of an entry page with a single search bar and a result page.
On the top of the result page is a search bar that allows for further text search.
Under the search bar is a row of textual intent elements detected by our language parsing model. 
To allow users to leverage the tag information available in some collections, for each detected element, an embedding-based search is performed across the tags in all the collections to find the most similar tags to each detected textual intent element.
Compared to existing tools like Google that only shows the relevant tags to the whole input, this design seeks to take advantage of the parsing to better organize the tag information and allow users to more transparently see which part of the input is related to which tags.
The main body of the result page displays the retrieved NFT items including the image and the collection and price information in a classic grid layout.
For close inspection of detail and visual parsing interaction, when users click on an individual image, a pop-up window is shown in the center of the page containing the interactable image and other interaction buttons to specify logical relations.

\subsection{Example Search Cases}
\begin{figure*}[t]
\centering
\includegraphics[width=0.995\textwidth]{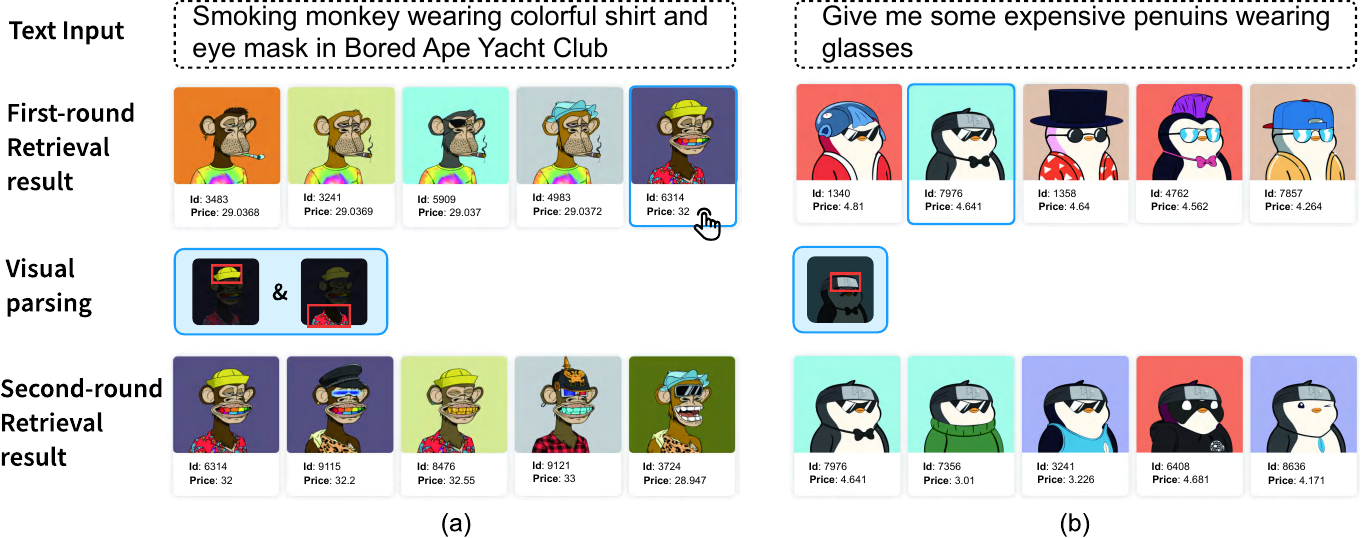}
\vspace{-2mm}
\caption{Examples of text search with visual parsing feedback. 
In the first case, the user inputs text to search for image with multiple visual features in the Bored Ape Yacht Club and then use visual parsing to find similar examples to an interesting result in top 5.
In the second case, the user searches for penguins with price requirement and finds an interesting ninjia headband he wants to focus on.
}
\label{fig:text_vis_search}
\end{figure*}
\textbf{Text search with visual parsing feedback}.
Figure~\ref{fig:text_vis_search} illustrates some examples of our intent expansion in search.
In the first case, the user inputs a rather long natural language expression containing multiple elements, including several elements about the visual content (monkey, smoking, colorful shirt, eye mask) and the collection metadata (Bored Ape Yacht Club).
With such complex queries, our language parsing can still understand quite well and find the best match in the top 3 return results.
But then the user may also find some unexpected results that seem to be quite rare and intriguing, such as the monkey wearing a yellow hat and a kind of red shirt with color patterns.
This time, the user switches to the visual parsing and utilizes its logic to select both the hat and the shirt to search for similar items with both visual elements.
In the results, the user can successfully find one more monkey with both elements.
In the second case in Figure~\ref{fig:text_vis_search}, the user inputs a query that contains both visual content (penguin, glasses) and a ranking preference for the price data (expensive).
The system returns some matching results and ranks them according to price from high to low.
Then, the user finds an interesting visual element that is somewhat like a ninjia headband.
So she uses visual parsing to continue exploring results with this element.

\begin{figure*}[t]
\centering
\includegraphics[width=0.995\textwidth]{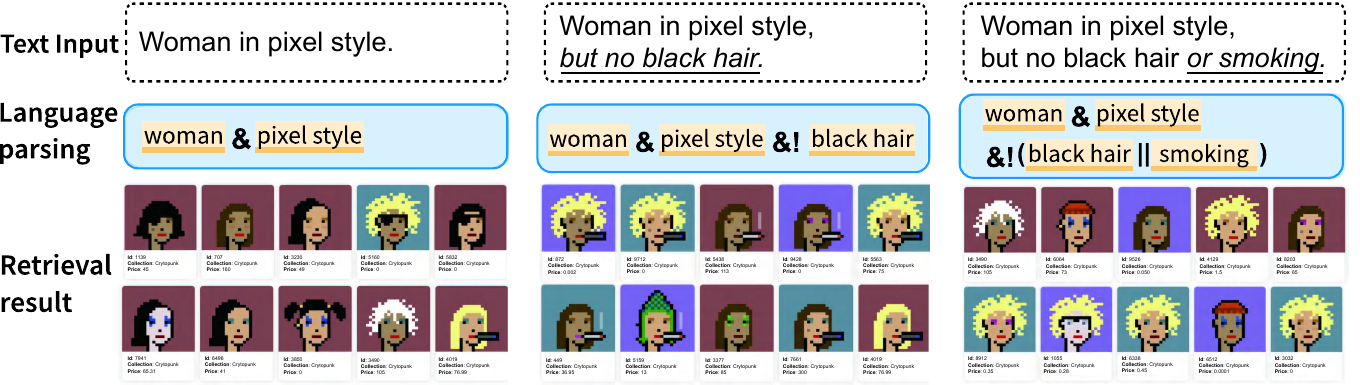}
\vspace{-2mm}
\caption{Example of logic understanding of text search. In the first case, the user searches for \emph{"woman in pixel style"}. In the second and third cases, the user adds the exclusion and union logic in natural language description to reduce unwanted visual elements, which are successfully understood by the system. }
\label{fig:text}
\end{figure*}
\textbf{Logic understanding of text search}.
Figure~\ref{fig:text} shows an example of the language logic composition capability of our framework.
The user first inputs \uinput{"woman in pixel style"} and gets some matching results.
But she finds that in the results there is too much black hair which she doesn't like.
So she adds the exclusion logic naturally at the end: \uinput{"but no black hair"}.
This successfully reduces black hair from results and only returns some yellow hair, brown hair and green hair woman.
However, she finds that this time there are too many NFTs that are smoking.
She further adds to the previous query and changes the single exclusion to a composition of exclusion and union: \uinput{"but no black hair or smoking"}.
The system can understand the complex logic in this search and return pixel style woman without black hair and cigarette.

\begin{figure*}[t]
\centering
\includegraphics[width=0.995\textwidth]{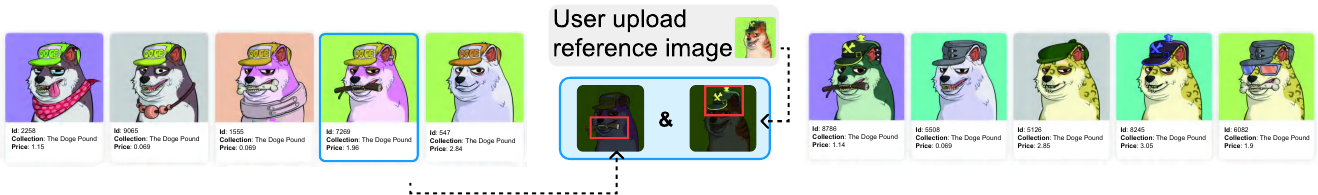}
\vspace{-2mm}
\caption{Example of visual logic composition. Users can not only select an element in one result image but also upload another image and select another element. Next they can logically compose the elements from the two different images to express a complex visual search intent.}
\label{fig:vis_compose}
\end{figure*}

\textbf{Visual logic composition}.
Figure~\ref{fig:vis_compose} shows an example of the visual logic composition.
The user finds an interesting piece of wood in a dog's mouth in the initial search results.
But she has also stored an image in previous search which contains a special cap she likes.
Thus, she wants to upload this previously stored image as a reference and selects the cap with visual parsing.
Then she performs the visual logic composed search with the intersection relation.
Finally, she finds a rather rare and interesting NFT image containing both elements she likes.

\begin{figure*}[t]
\centering
\includegraphics[width=0.995\textwidth]{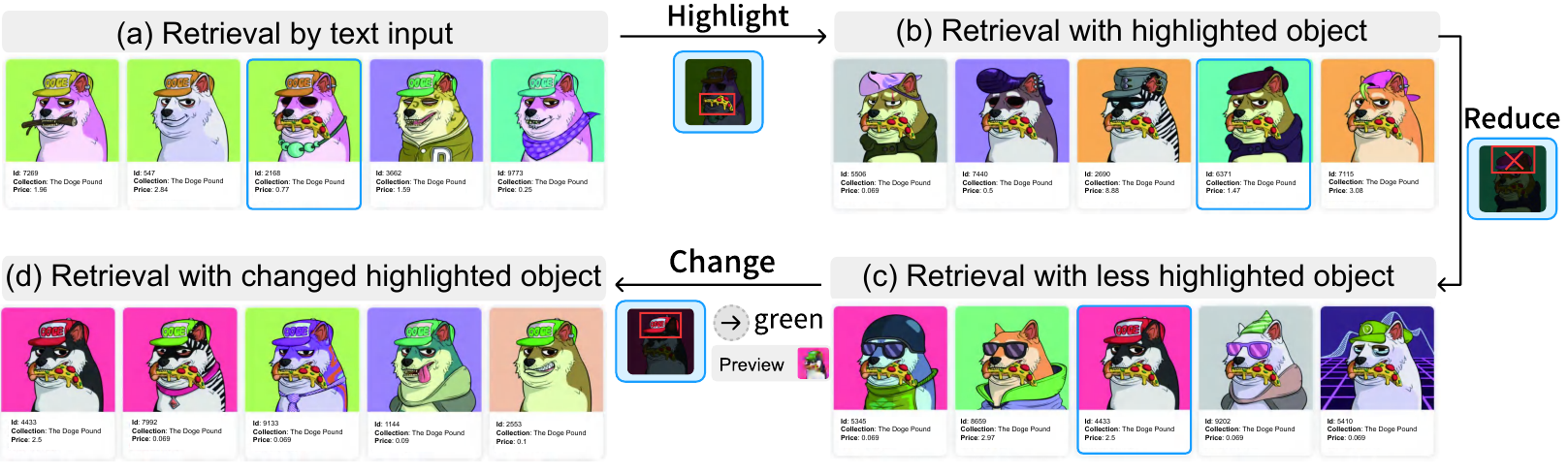}
\vspace{-2mm}
\caption{Examples of iterative exploration based on visual parsing and cross-modal interaction. User first sees some intial results (a) by text search, where she finds an interesting element-a piece of pizza in one result, which she selects and searches by visual parsing to get results in (b). She subsequently uses the exclusion logic to reduce a hat she does not like (c) and then uses the change logic to find similar works with an alternative hat color (d).}
\label{fig:vis_explore}
\end{figure*}
\textbf{Iterative search with visual parsing and cross-modal interaction}.
Figure~\ref{fig:vis_explore} shows an example of iterative use of our visual parsing in search.
In Figure~\ref{fig:vis_explore} (a), the user sees some initial results of dogs from text search.
She discovers one result image in which she finds an interesting element- a piece of pizza in the dog's mouth.
Then she leverages our visual parsing to quickly select the pizza and perform search, which gives her the results in Figure~\ref{fig:vis_explore} (b).
However, the user finds that there are two dogs in the results that wear a kind of hat she does not like.
She clicks on one of the dogs and selects the undesirable hat.
Then she exploits the exclusion logic in our visual parsing to reduce such hats from the results, with the new results shown in Figure~\ref{fig:vis_explore} (c).
Next, she finds one black dog with cap, pizza and red background she likes, but she wants to try changing the color of the dog's cap.
Therefore, she selects the cap and uses the change logic in our cross-modal logic composition to modify the element.
Before searching she also uses the preview function to see the effects of the modification and whether it is what she wants.
Finally, she performs the search and finds the most similar images according to her modified search intents in Figure~\ref{fig:vis_explore} (d).
\section{Evaluation}\label{sec:eval}

In this section, we report a within-subjects study to evaluate user perception and use of our system, compared to two existing search systems as the baselines, \ie, the unimodal text search used by many commercial applications and the cross-modal text-image search powered by recent advanced AI models like CLIP~\cite{radford2021learning}.   
Then, we focus on evaluating the new features of our intelligent parsing systems (\ie, language parsing, visual parsing, and logic compositions) and collect user satisfaction with these features.
\revise{Finally, we demonstrate the application of our method on a photo-realistic fashion image collection and provide objective quantitative metrics showing the benefit of our proposed framework.}

\subsection{Participant}
We recruited 16 participants aged between 21 and 35 (8 males, 8 females).
They are mostly graduate students studying digital media and art with diverse backgrounds, including computer science, engineering, art and design.
All of the participants have knowledge and experience about digital art and generative art and know about NFT.
Among the participants, six are developers specializing in generative art and design, five are designers and artists, and five are NFT researchers engaged in the study of NFT generation and recommendation.
\subsection{Baseline Search Systems}
In order to compare with existing search frameworks, we implement two alternative search systems for the same NFT data.
The first alternative is a system based on purely unimodal text search with metadata and tags.
Such systems represent many commercial image search platforms like Pinterest which rely on either the text accompanying the image on the webpage, the user-generated tags on the images or the object detection tags pre-generated by AI.
As this alternative method measures unimodal text similarity, we adopt the sentence transformer to produce text embeddings as they perform better than cross-modal embeddings in unimodal tasks. 
The second alternative is a system based on the latest cross-modal text-image search method which directly matches the text with the image through neural networks.
This system uses the same cross-modal CLIP embeddings as in our method, but it is not equipped with the intent expansion in our framework. 
The interface design remains the same for all the systems.
The two baselines are selected because they both allow for natural language input, which stands for most popular online image search platforms, such as Google, provides easier search interaction and is more comparable with our method.
Specifically, the unimodal text search baseline stands for most commercial systems like Pinterest, while the cross-modal text-image search baseline stands for the most recent advanced technology in AI-powered image search.

\subsection{Procedure and Data Analysis}
\textbf{Procedure.} In the study, we first introduced the background and purpose of our research.
Then, we introduced the functions of the baseline systems and the new features of our prototype.
Next, we let the participants freely explore our prototype and the baseline systems, where we asked them to perform five rounds of search in our system and the baseline systems.
The order of using different systems was randomly arranged.
Then, we asked the participants to complete a target search task by presenting them with 10 randomly sampled NFT images from our data.
They were asked to choose 1-3 images they like, use their language to describe the target image and search for similar results using both our system and the baseline system.
Next, the user was asked to try expressing more logic like the elements they don't like in the search.
Finally, the user was asked to try the visual parsing in our prototype to refine the results.
Upon completing the free exploration and targeted search tasks, we 
conducted a usability survey and semi-structured interview on our prototype in comparison to alternatives.
After the general usability survey and interview, we focused on our new features and asked participants to rate the usefulness of each.
Finally, we asked the user for further qualitative feedback and suggestions on our new features and the prototype in general.

\textbf{Analysis Methods.} We conducted mixed-methods analyses for users' quantitative and qualitative feedback. 
For quantitative analysis of the objective data, we measured the usability (ease of use), usefulness, integratedness, flexibility, interestingness, enjoyability and willingness to use in the future of our system and the baseline systems~\cite{brooke1996sus, yee2003faceted}.
Since most of the data did not follow a normal distribution, we carried on non-parameter tests.
Specifically, we ran two-way repeated measure ANOVA and posthoc analysis with Bonferroni correction for the quantitative data.
For qualitative analysis, two authors conducted an inductive thematic analysis of the open-ended feedback~\cite{hsieh2005three}. The final themes were established through iterative discussions and harmonization between the authors.

\subsection{Results}
\subsubsection{Prototype Usability}
To evaluate the usability of our prototype, we prepare the usability questionnaire based on our scenario~\cite{brooke1996sus, yee2003faceted}, combined with semi-structured interviews to gather further qualitative feedback.
The results show that our system scores relatively high ratings in all the metrics including ease of use ($mean_{our}=6.56, sd_{our}=0.61$; \revise{$mean_{uni}=5.19, sd_{uni}=1.55; mean_{cross}=6.19, sd_{cross}=0.88$}), usefulness ($mean_{our}=6.69, sd_{our}=0.58$; \revise{$mean_{uni}=4.69, sd_{uni}=0.85; mean_{cross}=5.75, sd_{cross}=0.97$}), integratedness ($mean_{our}=6.75, sd_{our}=0.56$; \revise{$mean_{uni}=3.94, sd_{uni}=0.90; mean_{cross}=5.06, sd_{cross}=1.09$}), flexibility ($mean_{our}=6.56, sd_{our}=0.50$; \revise{$mean_{uni}=4.13, sd_{uni}=1.17; mean_{cross}=5.63, sd_{cross}=0.93$}), interestingness ($mean_{our}=6.69, sd_{our}=0.46$; \revise{$mean_{uni}=4.13, sd_{uni}=0.60; mean_{cross}=5.50, sd_{cross}=0.61$}), enjoyability ($mean_{our}=6.88, sd_{our}=0.33$; \revise{$mean_{uni}=4.25, sd_{uni}=0.75; mean_{cross}=5.38, sd_{cross}=0.78$}), workload ($mean_{our}=6.25, sd_{our}=0.75$; \revise{$mean_{uni}=5.63, sd_{uni}=1.36; mean_{cross}=5.94, sd_{cross}=1.03$}) and future use ($mean_{our}=6.81, sd_{our}=0.39$; \revise{$mean_{uni}=3.94, sd_{uni}=1.20; mean_{cross}=5.31, sd_{cross}=1.21$}).

\begin{figure*}[t]

\centering
\includegraphics[width=0.995\textwidth]{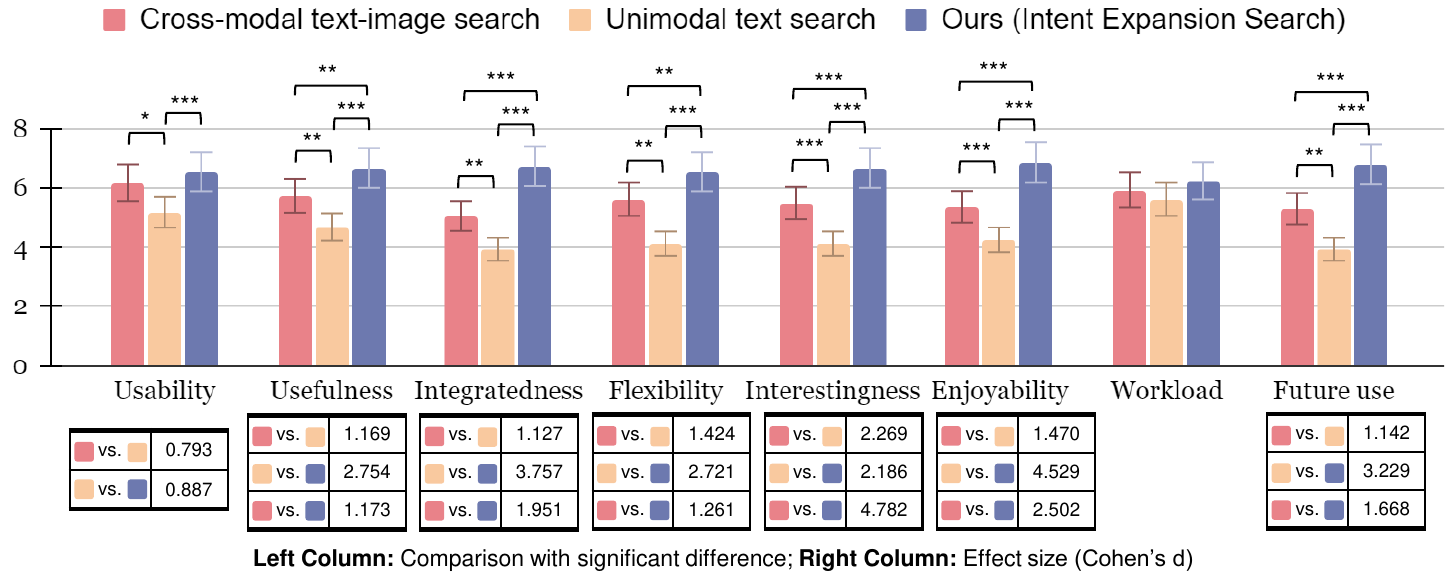}
\vspace{-2mm}
\caption{Usability Study. We evaluate the usability of our prototype compared with baseline systems on the same dataset. 
% The usability metrics range from easiness of use to interestingness of the search experience. The baseline systems are the unimodal text search and the cross-modal text-image search.
The results show that our prototype generally scores higher across the usability metrics than the baselines, with significant differences in usefulness, integratedness, flexibility, interestingness, enjoyability and willing to use in the future.}
\label{fig:user1}
\end{figure*}

For the significant score, a two-way repeated ANOVA revealed that participants perceived our system significantly better than the other two baselines in the quantitative metrics except for the workload.
Specifically, participants mostly considered our system easier to use than baselines ($p<0.001$).
They also considered our search system more useful ($p<0.001$) than the others ($p<0.001$).
In terms of integratedness, which refers to the integration of different types of input, data features, and search functions,  they also mostly prefered our system ($p<0.001$).
In addition, they thought our system support more flexible search ($p<0.001$).
The search experience in our system is also more interesting ($p<0.001$) and enjoyable ($p<0.001$).
Finally, they all expressed more willingness to use our system than the baselines in the future ($p<0.001$).
The significance scores \revise{and Cohen's d as an effect size indicator for significant comparisons} are shown in Figure~\ref{fig:user1}.

\begin{figure*}[t]

\centering
\includegraphics[width=0.995\textwidth]{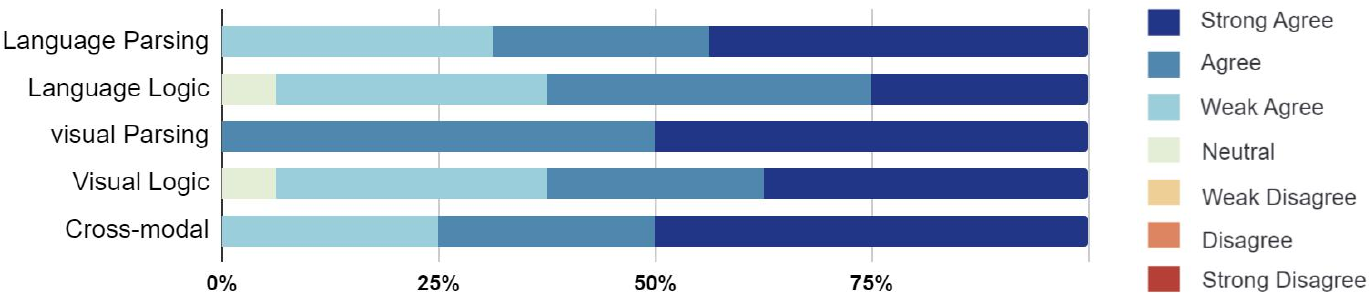}
\vspace{-2mm}
\caption{User evaluation of the usefulness of our unique new functions. 
Visual parsing is deemed most useful, while participants tend to rate the parsing slightly higher than the logic composition.
}
\label{fig:func}
\end{figure*}

\subsubsection{Qualitative Feedback for Survey}
\begin{itemize}
    \item 
    \textbf{Ease of use}.
     Participants reported that our prototype is easier to use because it is easier to find proper visual input for image-to-image search (5/16) compared to the baselines.
     As P7 said, \emph{"It is very difficult for me to find example images that match exactly what I want to search in other systems, but I can use this system (our)'s feedback to identify specific visual features I want more easily."}
     In addition, most participants (14/16) felt that our visual parsing is intuitive and easy to use.
     For example, P2 said, \emph{"Clicking and brushing is a very easy interaction I have been accustomed to in traditional systems; it does not require any extra complex interaction and takes only about a second to finish. But the parsing result is impressive."}
     In comparison, the unimodal baseline was considered the most difficult to use because all participants found that it is too sensitive to the keyword they entered.
     However, some participants (4/16) found that the language parsing occasionally misunderstood their price requirements in the search due to the uncertainty of GPT and they needed to slightly alter their inputs. 
    \item 
    \textbf{Usefulness}.
    Many participants thought our system is more useful compared to cross-modal and unimodal baselines.
    % This is mainly because 
    First, they think visual parsing which helps them focus on specific visual elements is very useful as it provides a more accurate visual search (9/16).
    \emph{"When I want to see more results with this specific type of flower decoration, it can help me focus on this element and return more matching results."}, P6 mentioned.
    Second, they can search for more combinations of features in our system (7/16).
    For example, P14 commented, \emph{"This system (our) seems to understand the multiple keywords I input slightly better than the other two. And I can also select multiple visual elements here."}
    Third, language parsing is useful for improving search accuracy and incorporating metadata (5/16).

    \item
    \textbf{Integratedness}.
    All participants perceived our system is more integrated compared with baselines.
    Language parsing allows them to search different information in one input such as image description plus the price description (12/16).
    For example, P5 said, \emph{"It is indeed convenient to be able to search more information in a single search bar without additional operations, particularly for NFT I think the price data is often important as well."}
    The cross-modal logic and interactions also integrate text and image input (6/16).
    For example, P1 remarked, \emph{"The composition of text and image allows me to describe more complex requirements such as editing some features, which is not possible with usual text or image input."}
    However,  a few participants (2/16) commented that even though our preview function is interesting, it seemed to be a stand-alone function only for the text-guided change logic.
    \emph{"It would be nice if I could preview the effects of other compositions such as exclusion,"} P12 commented.

    \item 
    \textbf{Flexibility}.
    More than half of the participants said our system is more flexible than the cross-modal and unimodal baselines.
    Because our system simultaneously supports enhanced user intent expression in text, image, and cross-modal input as well as feedback, participants feel they have more choices for search input and interaction (9/16).
    For example, P11 said, \emph{"I could switch between natural language expression and intuitive visual input. When the results of both are not satisfactory, I also tried the cross-modal inputs. In any case, because none of the image search systems have perfect matching, I can find more relevant results with more diverse inputs."}
    However, some participants (3/16) noted that sometimes they could not flexibly adjust the visual parsing results in cases they were not satisfied with the parsing, as they could only clear the parsing and start from scratch.

    \item 
    \textbf{Interestingness}.
    Most participants said our system is more interesting than cross-modal and unimodal baselines.
    This is because our visual parsing and cross-modal interactions are a novel experience to them (13/16).
    For example, P14 said \emph{"it is really interesting to play with the result images with brushing and editing. This reminds me of my experience with creative tools like Photoshop."}

    \item 
    \textbf{Enjoyability}.
    Similar to the interestingness, most participants (14/16) said the search experience provided by our system is more enjoyable than the baselines, mainly for the reason that our interactive feedback provides a more enjoyable iterative search experience.
    For example, P12 said \emph{"I would like to iterate the search with the feedback multiple times in this system (ours) and really enjoy the process to dig deeper into the image gallery. I haven't had much experience with any other image search systems I have used."}

    \item 
    \textbf{Future Use}.
    All the participants would like to use our system in the future because it generally offers a more accurate search and better user experience than the existing systems \revise{they} have used (14/16).
    For example, P1 said, \emph{"The image search systems I have used before feel rather old-fashioned and I thought research in image retrieval is quite boring. But when I see you integrated cross-modal interactions and even some generation mechanisms, I think it shows more potential for modern image search systems, especially in the age when multi-modal data floods the Internet."}

\end{itemize}

\subsubsection{Search Functions Evaluation}
We also conducted a 7-Likert-scale survey on the usefulness of each unique function in our prototype, including language parsing and logic, visual parsing and logic as well as cross-modal interaction.
As shown in Figure~\ref{fig:func}, the results show that most participants think the functions of our prototype are useful.

In general, most participants acknowledged the usefulness of our functions.
Particularly they mostly appreciate the visual parsing function ($mean=6.50, sd=0.50$) more than other functions.
This is because visual parsing is both impressive to them and useful for improving the accuracy of iterative visual search.
However, a few participants do not particularly appreciate the logic.
For example, P16 said sometimes the system has some errors and does not understand some logic expressed in the input.
On average participants give higher ratings on the parsing than the logic in both language and visual modalities.
The reason is that they think parsing is the foundation of improving the system's ability to capture specific elements in user intent, while the logic is dependent on the parsing and seeks to provide more accurate fine-grained operation.
\emph{"The visual logic is useful in some cases but I suppose I would not use it as often as the visual parsing itself as sometimes I only want to focus on a single element,"} P12 commented.
Participants are also mostly positive about the usefulness of our cross-modal interactions ($mean=6.25, sd=0.83$).
However, some of them also suggest that the text-guided masked editing of images is too stringent.
As P14 said, \emph{"I hope I could use a more free editing model or even unload my own fine-tuned model weights to customize the editing and the preview."}

\subsection{More Insights from User Feedback}
\label{ssec:more_feedback}
We summarize some additional interesting insights that could inspire a more user-friendly image search design according to our interview with the participants.

\textbf{Incorporation of easy-to-use feedback interaction}.
Some participants (3/16) particularly noted that the easiness of interaction is an important strength of our system and the design of feedback should not get too complex.
This means that there is a trade-off between fine-grained interaction and usability.
As P4 commented, \emph{"I would be less inclined to use the feedback if I need to perform more complex input action like sketching or typing in long additional text."}
However, some participants also suggest other types of easy-to-use interactions they are familiar with in non-retrieval systems to enrich the user experience. 
For example, P9 said, \emph{"Brushing by box selection is easy to use and quite accurate but I also want a lasso selection as I'm familiar with it in authoring tool even though it requires more complex interaction."}
P7 commented, \emph{"The interactive feedback in this system (our) is really nice. 
But I also suggest other intuitive interactions; you can take inspiration from a recommendation system and allow users to instantly remove a single image from the retrieved results and record user preferences implicitly for subsequent search."}

\textbf{Compositionality and other aesthetics}.
Some participants (5/16) said our parsing-based search is particularly suitable for exploring artworks with composition as an important factor.
P6 said, \emph{"NFT is a type of artwork where the composition of elements is evident and plays an essential role in its aesthetics. Even the evaluation of an NFT's rarity and its potential value can be linked to such features. So I think the parsing is particularly useful for NFT."}
However, \emph{"Other types of artworks with more complex features have additional compositional features including spatial layout, golden ratio or geometric shape of the outline, which this system (ours) still cannot support."}
In addition, some artists (P12, P14) also expect further incorporation of aesthetic information in the search, such as detailed specifications of color palette.

\textbf{Transparency and controllability}.
Some participants (6/16) said the intent recognition of text input can increase the transparency of the AI behind the search as the intent elements are displayed and they can see how AI understands their natural language input.
However, some participants (3/16) would like further control through more complex interactions, because in some cases they may not be satisfied with the system's interpretation of their intents and want to adjust it in even more customized manner.
For example, P8 said, \emph{"I think additional interaction that can help me adjust the weights of multiple elements to influence the retrieval results or even use the special mark on a particular text element to stress my intent to prioritize it."}
P9 and P13 said they would like more control over the visual parsing with interaction tools such as an eraser to clear part of the accidentally selected region.

\textbf{The relationship between generation and retrieval}.
Participants noted that with the rapid development of AI generation models, image generation and image retrieval are no longer two independent tasks.
Some participants (6/16) commented that our preview function in the composed visual feedback supported by the generative editing model enables them to see alternative results.
In cases when they cannot find exactly matching examples in the database, they can at least see examples from the edited result.
This can serve as a visual aid for them to inspect or confirm whether the modified result is really what they want in terms of aesthetics.
However, some participants (3/16) also pointed out that they would like to integrate more generation functions in the system, particularly for NFT.
As P1 said, \emph{"I'm particularly interested in using powerful AI to generate my own NFT. However, because the community is very important for the publicity and trade of NFT, I would like to imitate the style of the popular collection instead of my own. So, I wish I can directly use the retrieval results as references for AI to generate more diverse variations and check for its rarity in the marketplace. This system (ours) can support simple modification of selected elements, but I would like more flexible generation allowing more sophisticated generation prompt."}
In addition, P8 pointed out that the preview function can be better integrated into our search system: \emph{"I think the preview is interesting as it somewhat helps me visualize my modified search intents. However, such a function can also be integrated with the initial pure text search to make the whole interaction design more consistent, as for the initial text search, such visualization of users' search intent is also helpful for users to visually confirm what they want, especially when they are looking for references for designing their own work."}

\subsection{Evaluation on photo-realistic benchmark dataset}
\label{ssec:photo}

\revise{In order to provide some quantitative evidence for the benefit of our method, and show the potential applicability of our framework to other styles of image data, we test our method on a benchmark photo-realistic image dataset, Deepfashion2~\cite{ge2019deepfashion2}, as no benchmark dataset for NFT data is available.
Deepfashion2 is a large dataset with commercial and customer fashion images.
We choose its validation set containing 32,153 images to build the retrieval database.
Then, we synthesize 500 queries by sampling both image and annotation data from Deepfashion2's training set, which does not have overlapping images with the validation set.
Subsequently, we evaluate our method's ability to improve understanding of users' text and image input by comparing traditional input with our intent expansion (IE) enhanced method.
For fair comparison, all the methods use the original CLIP model as the text and image encoders, and no metadata in the validation database is used for the retrieval.
We use the Top-K accuracy to measure the performance. 
From the results in Table~\ref{tab:quant}, we can see that our methods can indeed consistently improve the retrieval results in terms of accuracy.
However, we also notice that for photo-realistic images with more complex features than NFT images, the retrieval results still leave large room for improvement.
This is partly because the foundational CLIP model still does not have adequate understanding of photo-realistic fashion images, which requires more domain metadata to finetune, but is beyond the focus of our study.
}

\begin{table}
\centering
\begin{tabular}{|c|c|c|c|c|}
& \multicolumn{2}{c|}{\textbf{Image Input}} &  \multicolumn{2}{c|}{\textbf{Text Input}} \\\hline
      & without IE & \textbf{ours} & without IE & \textbf{ours}\\\hline
\textbf{Top-1} & 39.80\% &\textbf{42.80\%} & 37.00\%& \textbf{60.80\%}\\\hline
\textbf{Top-5} & 39.92\%&\textbf{41.32\%} & 41.76\%& \textbf{46.52\%}\\\hline
\textbf{Top-20}& 36.65\%& \textbf{40.41\%} & 45.05\% & \textbf{45.76\%} \\
\end{tabular}
\caption{\revise{Quantitative evaluation with Top-K metrics of the effect of our intent expansion (IE) on improving the CLIP-based retrieval for photo-realistic fashion data.}}
\label{tab:quant}
\end{table}

\begin{figure*}[t]

\centering
\includegraphics[width=0.995\textwidth]{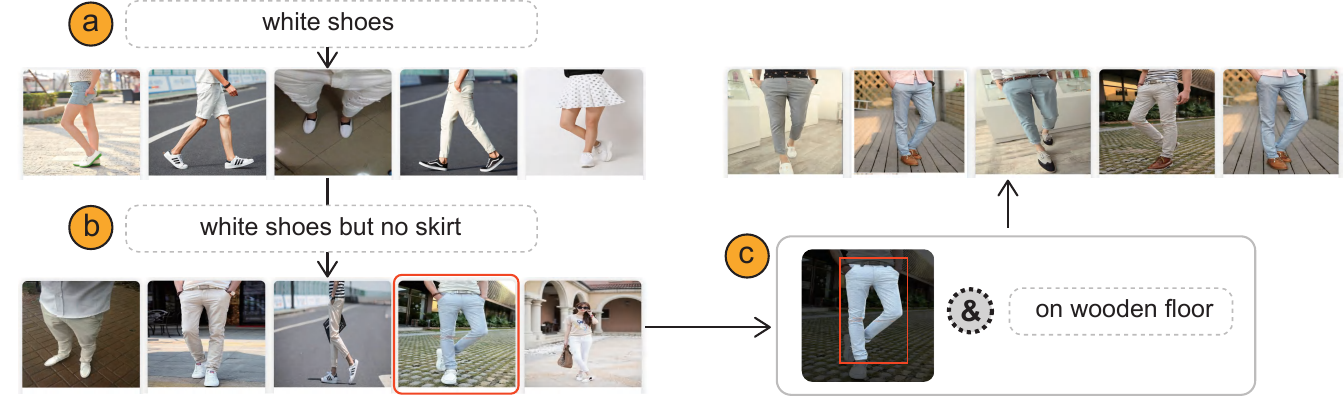}
\vspace{-2mm}
\caption{\revise{An example case of using our method on photo-realistic fashion image data. 
(a) User first inputs a short phrase white shoes and retrieve some relevant images.
(b) Then, the user can add logic to the natural text input to exclude images with skirt.
(c) Next, the user can select an image with blue jeans and combine visual parsing and multi-modal logic to retrieve some images with similar jeans and wooden floor.}
}

\label{fig:fashion}
\end{figure*}
\revise{As shown in Figure~\ref{fig:fashion}, we also provide an example illustrating the application of our method to photo-realistic fashion images without using metadata.
In this example, the user first uses text search, which is assisted by our language parsing and logic to understand the exclusion intention, as in Figure~\ref{fig:fashion} (a) and (b).
Then, the user finds an interesting pair of jeans from the results.
He or she can combine visual parsing with multi-modal logic composition to find images of models wearing similar jeans and standing on wooden floor to see how it fits in such environment.
}

\section{Discussion}\label{sec:discuss}
\noindent
\textbf{Further improvement of visual content match}. Some users noted that the visual content matching may require further improvement for specific domains and styles because they can observe in both the cross-modal baseline and our prototype that there are some misunderstandings of image content sometimes, and more advanced computer vision techniques need to be developed and incorporated into the modern image search system.

\noindent
\textbf{Multilingual support}. Because we build our prototype for the scenario of image search in the NFT domain which mostly concerns metadata and tags in English, we only focus on English input.
In addition, some of the fine-grained multi-modal interaction such as the modification and preview relies on AI models which are mostly trained in the English language.
But some users mentioned that there are also large galleries featuring culture-specific images such as traditional Chinese paintings that can be best described in the Chinese language.
P7 even said, \emph{"What I want is a system that can simultaneously accept input of multiple languages."}

\noindent
\textbf{Broader applications}. Some participants (P12-P14) suggest that our framework can be applied to art galleries of different styles. 
For example, the current digital image galleries for traditional paintings such as WikiArt and the Web Gallery of Art still mainly rely on traditional metadata or tag-based search.
For paintings, it is also important to retrieve multi-modal information including its essential visual features and metadata about its historical context.
In addition, fine-grained visual parsing can be leveraged to perform more detailed and intuitive aesthetic-aware retrieval.
Another future application domain is the latest AI-generated images.
Many tools and platforms like Midjourney and Stable Diffusion WebUI enable the fast and easy generation of large numbers of images with important text prompt metadata.
Users need efficient exploration \revise{for} the huge and ever-growing number of generated images shared by other AI users.
\revise{In addition, our approach has the potential to be applied to real life photo-realistic images, as we show in Section~\ref{ssec:photo}.
However, without high-quality metadata, the complex features in photo-realistic images may affect user experience as they pose more challenges for visual parsing and multi-modal logical composition due to factors like less clear-cut visual elements with more irregular shapes and textures.
The multi-modal alignment for foundational model like CLIP can also be challenging for photo-realistic data from diverse domains which cover different objects and require varying levels of detail.
For example, a large and messy indoor scene is much more complex than a fashion image which only contains a subject and a few garments, making it more difficult to distinguish small elements from each other.}

\noindent
\textbf{Other modalities}.
Our study mainly focuses on textual and visual input for image retrieval.
However, our framework can be enriched to incorporate other modalities.
For example, language parsing can be enhanced by combination with speech recognition to allow users to provide audio input.
In addition, many media data, such as some special collections of NFT are in other formats like video.
Our method can be extended to video for a better search experience.
In this case, more complex parsing involving temporal information needs to be developed.

\noindent
\textbf{More flexible visual cues or interactions for visual parsing}.
In the visual parsing of our framework, we only implement the box selection based on rectangle brushing.
Even though such interaction is quick and easy to use, additional interactions can be incorporated to further enhance the experience.
For example, occasionally our AI-driven semantic parsing may misunderstand user intents and select a larger semantic region than expected.
In such cases, there is no way to correct it except by clearing the mask and starting over.
To address such issues, some interactions such as users adding a negative point on the image to mark the neighborhood to be excluded~\cite{kirillov2023segment} can facilitate more customized user control.
Such point-based interaction can be also used to express positive selection intents~\cite{kirillov2023segment}, but such interactions often require marking out multiple points.
More experiments need to be done to evaluate how such interactions can benefit the contextualized feedback of image retrieval.

\noindent
\textbf{Content creation and retrieval}.
As we discussed in the user feedback in Section~\ref{ssec:more_feedback}, some users envision more integration of retrieval and generation in web image galleries.
Traditional web image galleries have focused primarily on retrieval, allowing users to search and browse through a collection of images based on specific criteria such as tags, categories, or keywords.
With the advancement of generative models such as deep learning-based image synthesis techniques, there is potential for web image galleries to generate entirely new images based on user-defined criteria. 
Users could input textual descriptions or even rough sketches, and the gallery could generate images that match the given specifications. 
This integration will open up new possibilities for creative expression and exploration within web image galleries.
However, the integration of retrieval and generation could also potentially cause problems.
One problem is the ethical or copyright issues.
Many AI-powered image generation models allow users to input reference images they retrieve from the web to create variations based on the initial image.
Many AI users would consider such generated images their own creation and upload them to web galleries under their names.
This can potentially infringe on the rights of the original artists.
Even worse, it could discourage original artists from publishing their works in web galleries if the images are easily retrievable by AI users to generate their own works.

\section{Future Work}
\textbf{Enrich intent expansion to support digital assets search in different domains and modalities}.
In future work, we would apply our framework to wider range of domains including art and design image galleries of different styles, as well as other digital assets such as images, videos or even 3D models of cultural heritage.
For these applications in new domains and new modalities, richer interactions and more detailed user intent expansion need to be developed.
For example, for searching of 3D digital assets, in order to allow users to perform contextualized interactions, we need to incorporate efficient visual parsing or spatial parsing in 3D space. 
We also need 3D editing AI technology to allow users to express their modification intents and preview the results.
For other artistic images, we need to allow users to express more fine-grained aesthetic intents, such as associating their words with specific color palette or specific compositional patterns.

\noindent
\textbf{Increase controllability of intent expansion}.
Currently we rely on AI-powered vision-language models as mediators to enhance understandingg of users' search intents.
However, as shown in Section~\ref{sec:eval}, the AI models would occasionally misunderstand user inputs either in language or visual parsing.
Some users also express their preference for more control on the intermediate process.
Thus, in future work, we would build on the intent expansion by incorporating user control of AI's parsing and composition of linguistic or visual elements. 
For example, we can leverage users' real time feedback to finetune the language parsing or visual parsing model.
We can also allow users to specify the weights of different elements in composition.
\section{Conclusion}\label{sec:conclude}

This paper presents an intent expansion framework that aims to enhance the image search experience. The framework is designed based on user requirements gathered through a formative study, which emphasizes the need for a better understanding of users' search intents and improving user feedback.
The intent expansion framework consists of two main components: intent parsing and logic composition. The intent parsing module utilizes vision-language models to recognize detailed search intents from both textual and visual inputs provided by the users. This helps in capturing the users' specific requirements more accurately.
The logic composition component leverages chain-of-thought prompting and user interaction to compose intent elements expressed in both text and visual input. By facilitating the composed search of multiple elements, this component enables users to conduct more sophisticated and targeted searches.
Additionally, the intent expansion framework supports contextualized cross-modal interactions, allowing users to iteratively refine their searches based on the initial search results. 
This iterative process improves the overall search experience by providing users with relevant and tailored results based on their interactions with the system.
We build a prototype for the scenario of NFT image search based on the framework.
Then we conduct a user study to evaluate the \revise{system's} usability and the new functions of the framework.
The NFT data we use is available at \url{https://osf.io/hkm6b/}.

\section*{Acknowledgment}
The authors wish to thank the anonymous reviewers for their valuable comments.
This work was supported by the National Natural Science Foundation of China (No. 62172398) and Guangzhou Basic and Applied Basic Research Foundation (2023A03J00059).

\bibliographystyle{ACM-Reference-Format}
\bibliography{sample-base}

\end{document}